\documentclass[12pt]{article}

\usepackage{amsmath}
\usepackage{amsfonts}
\usepackage{cite}
\usepackage{graphicx}

\numberwithin{equation}{section}

\def\beq#1\eeq{\begin{equation}#1\end{equation}}
\def\bes#1\ees{\begin{equation}\begin{split}#1
               \end{split}\end{equation}}
\def\bea#1\eea{\begin{align}#1\end{align}}

\newcommand{\abs}[1]{\lvert #1 \rvert}
\newcommand{\norm}[1]{\lVert #1 \rVert}

\newcommand{\Z}{\mathbb{Z}}
\newcommand{\Q}{\mathbb{Q}}
\newcommand{\R}{\mathbb{R}}

\newcommand{\bra}[1]{\langle #1|}
\newcommand{\ket}[1]{|#1\rangle}

\newcommand{\dbra}[1]{\langle\!\langle #1|}
\newcommand{\dket}[1]{|#1 \rangle\!\rangle}

\newcommand{\OO}[1]{\mathcal{O}(#1)}
\newcommand{\auto}[1]{G(#1)}
\newcommand{\fusion}[4][\mathcal{N}]{{#1}_{#2 #3}{}^{#4}}

\newcommand{\A}{\mathcal{A}}
\newcommand{\I}{\mathcal{I}}
\newcommand{\E}{\mathcal{E}}
\newcommand{\V}{\mathcal{V}}

\newcommand{\G}{G_\mathrm{sc}}          
\newcommand{\Gid}{G_\mathrm{id}}        
\newcommand{\stab}[1]{\mathcal{S}(#1)}  
\newcommand{\stabc}[1]{\mathcal{S}^c(#1)}  

\DeclareMathOperator*{\dirsum}{\oplus}

\begin{document} 
\baselineskip=6mm

\begin{titlepage}
\nopagebreak
\vskip 5mm
\begin{flushright}
hep-th/0207177\\
TU-664
\end{flushright}

\vskip 10mm
\begin{center}
{\Large \textbf{Novel construction of boundary states in 
coset conformal field theories}}
\vskip 15mm
Hiroshi \textsc{Ishikawa}
\footnote{\tt ishikawa@tuhep.phys.tohoku.ac.jp} 
and Taro \textsc{Tani}
\footnote{\tt tani@tuhep.phys.tohoku.ac.jp}
\vskip 5mm
\textsl{%
Department of Physics, Tohoku University \\
Sendai 980-8578, JAPAN\\}
\end{center}
\vskip 15mm

\begin{quote}
We develop a systematic method to solve the Cardy condition
for the coset conformal field theory $G/H$.
The problem is equivalent to finding a non-negative integer
valued matrix representation (NIM-rep) of the fusion algebra.
Based on the relation of the $G/H$ theory with the tensor product theory
$G \times H$,
we give a map from NIM-reps of $G \times H$ to those of $G/H$. 
Our map provides a large class of NIM-reps in coset theories.
In particular, we give some examples of NIM-reps
not factorizable into the $G$ and the $H$ sectors. 
The action of the simple currents on NIM-reps
plays an essential role in our construction. 
As an illustration of our procedure,
we consider the diagonal coset $SU(2)_5 \times SU(2)_3 /SU(2)_8$
to obtain a new NIM-rep based on the conformal embedding
$su(2)_3 \oplus su(2)_8 \subset sp(6)_1$.
\end{quote}

\vfill
\end{titlepage}

\section{Introduction}

The study of the boundary conditions in rational conformal
field theories (RCFTs) has been attracted much attention
since the pioneering work by Cardy \cite{Cardy}.
It is now recognized that 
the requirement of the mutual consistency of the boundary
conditions, which is known as the sewing relations
\cite{CardyLewellen,Lewellen}, 
is so severe that the number of the possible
solutions is finite if we keep the chiral algebra of RCFT at 
the boundaries
\cite{Cardy,CardyLewellen,Lewellen,Pradisi,Runkel,%
FS, BFS, BPPZ,FFFS,BS}.
The Cardy condition \cite{Cardy} is one of the sewing relations,
which expresses the consistency of the annulus amplitudes without
insertion of vertex operators.
The problem of solving the Cardy condition is,
under the assumption of the completeness \cite{Pradisi},
reduced to the study of the non-negative integer valued matrix
representation (NIM-rep) of the fusion algebra\cite{BPPZ,DZ,Gannon}.

In the context of string theory, 
the existence of the boundaries on the worldsheet means
the presence of $D$-branes. 
Since the (super) conformal invariance on the worldsheet is necessary 
for the consistency of string theory,
the possible boundary conditions are those preserving the conformal
invariance and we have boundary CFTs. 
The classification of the possible $D$-branes is therefore
equivalent with the classification of the conformal boundary conditions
in a given CFT. 
Although RCFTs describe only some specific points in the moduli space
of the string backgrounds, 
we can obtain insights into the stringy nature of $D$-branes
from the study of the boundary conditions in RCFTs.

The coset CFT \cite{GKO} is an important class of RCFTs,
which contains in particular the minimal models. 
The coset theories are considered to be the building blocks of
generic RCFTs. 
Therefore it is of fundamental importance to investigate
and classify the boundary conditions allowed in the coset theories.
This problem has been studied by several groups
\cite{MMS,Gawedzki,ES,Fredenhagen,Ishikawa,Kubota,Nozaki,Quella,%
Fredenhagen2,Walton}.
The issue of the classification is however not completely answered yet.

In this paper, we start a systematic study toward 
the classification of the boundary conditions in the coset theories. 
Based on the relation of the coset theory $G/H$ 
with the tensor product theory $G \times H$, 
we show that any NIM-rep of the $G \times H$ theory yields
a NIM-rep of the $G/H$ theory.
This map from $G \times H$ to $G/H$ enables us to construct
a large class of NIM-reps in the $G/H$ theory.
In particular, we can obtain some examples of NIM-reps 
not factorizable into the $G$ and the $H$ sectors. 
The construction in this paper generalizes that given in 
the previous work by one of the present authors \cite{Ishikawa},
in which the resulting NIM-reps of the $G/H$ theory
are only of the factorizable form. 
Our map is nothing but the NIM-rep version
of the map for the modular invariants of 
the $G \times H$ and the $G/H$ theories obtained in \cite{GannonWalton}.

The action of the simple currents
\cite{SY,Intriligator} on generic NIM-reps
\cite{Ishikawa,GG} plays an essential role in our construction.
Namely, it is shown that
we need appropriate identifications
and selection rules for the Cardy states of $G \times H$,
which are generated by the action of the simple currents
on the Cardy states,
to obtain a NIM-rep of $G/H$. 
These brane identifications and brane selection rules
are the counterparts of the field identification and the selection rule
of the coset theories, and has been first observed in \cite{Ishikawa}.

The organization of this paper is as follows.
In the next section, after reviewing some basic facts about
the boundary states in RCFTs (Section \ref{sec:NIM}),
we develop the tools necessary in our construction
of the NIM-reps in coset theories. 
First, we argue the action of the simple currents on generic NIM-reps
in RCFTs and give the definition of the simple current group
for NIM-reps (Section \ref{sec:sconNIM}).
Next, we explain methods to construct a NIM-rep
from a given one: the use of the simple currents and
conformal embeddings (Section \ref{sec:NIMconstruction}).
In Section \ref{sec:tensor}, 
we discuss the NIM-reps in tensor product theories.
Applying the methods developed in Section \ref{sec:NIMconstruction}, 
we give some nontrivial NIM-reps in the $SU(2)_k \times SU(2)_l$ theory. 
In Section \ref{sec:coset}, we turn to coset theories and
give a map for NIM-reps of the $G \times H$ and the $G/H$ theories. 
The necessary identifications and selection rules for
the Cardy states of the $G \times H$ theory is described in terms of
the simple current groups for NIM-reps obtained in Section \ref{sec:sconNIM}
and the field identification currents of the coset theory. 
As an illustrative example for our procedure,
we consider the diagonal coset $SU(2)_k \times SU(2)_l /SU(2)_{k+l}$
to obtain a NIM-rep not factorizable into the numerator and
the denominator sectors.
The final section is devoted to some discussions.


\section{Boundary states in rational conformal field theories}

\subsection{NIM-reps in RCFTs}
\label{sec:NIM}

In this subsection, we review some results on 
the boundary states in rational conformal field theories (RCFTs).

The most important ingredient of a RCFT is a chiral algebra
$\A$, which is the Virasoro algebra or an extension thereof.
We denote by $\I$ the set of 
all the possible irreducible representations of $\A$.
For a RCFT, $\I$ is a finite set.
In the bulk theory, we have a pair of algebras $\A$ and $\tilde{\A}$,
which correspond to the holomorphic and the anti-holomorphic
sectors, respectively.
In the presence of boundaries, $\A$ and $\tilde{\A}$ are related with
each other via appropriate boundary conditions.
We restrict ourselves to the case of rational boundary conditions,
which preserve the chiral algebra $\A$ at the boundary.
Since $\A$ includes the Virasoro algebra, the rational boundary 
condition keeps the conformal invariance at the boundary
and yields a boundary CFT.

A boundary state $\ket{\alpha}$ is characterized by the following
equation
\footnote{%
One can twist the boundary condition with an automorphism $\omega$ of
$\A$ to keep the rationality of the theory.
In this paper, however, we consider the case of the
trivial automorphism $\omega = 1$.}
\beq
   (W_n - (-1)^h \widetilde{W}_{-n}) \ket{\alpha} =0 . 
\label{eq:symmetricbc}
\eeq
Here $W$ and $\widetilde{W}$ are the currents of 
$\A$ and $\tilde{\A}$, respectively.
$h$ is the conformal dimension of the currents:
$h=2$ for the Virasoro algebra and $h=1$ for the affine Lie
algebras.
The building blocks of the boundary states are 
the Ishibashi states $\dket{\mu}$ ($\mu \in \I$) \cite{Ishibashi}.
We normalize the Ishibashi states as follows,
\beq
   \dbra{\mu} \tilde{q}^{H_c} \dket{\mu'}              
   = \delta_{\mu\mu'}\frac{1}{S_{0\mu}}\chi_{\mu}(-1/\tau)
   = \delta_{\mu\mu'}
     \sum_{\nu \in \I}
        \frac{S_{\mu\nu}}{S_{0\mu}}
        \chi_{\nu}(\tau)
  = \delta_{\mu\mu'}(\chi_0(\tau) + \cdots ),
\label{eq:Ishibashinorm}
\eeq
where $\tilde{q} = e^{-2\pi i /\tau}$.
$H_c = \frac{1}{2}(L_0 + \tilde{L}_0 - \frac{c}{12})$ 
is the closed string Hamiltonian and
$c$ is the central charge of the theory.
$\chi_\mu$ is the character of the representation $\mu \in \I$,
and its modular transformation reads
\beq
 \chi_{\mu}(-1/\tau)= \sum_{\nu \in \I}
                      S_{\mu\nu}\chi_{\nu}(\tau).
\label{eq:Smatrix}
\eeq
`$0$' stands for the vacuum representation.
The normalization \eqref{eq:Ishibashinorm}
corresponds to the following scalar product 
in the space of the boundary states \cite{BPPZ}
\beq
  \bra{\alpha} \ket{\beta} =
  \lim_{q \rightarrow 0} q^{\frac{c}{24}} 
  \bra{\alpha} \tilde{q}^{H_c} \ket{\beta} ,
\eeq
where $q = e^{2\pi i \tau}$.

We denote by $\V$ the set labeling the boundary states. 
A generic boundary state $\ket{\alpha} (\alpha \in \V)$ 
satisfying the boundary condition \eqref{eq:symmetricbc} is 
a linear combination of the Ishibashi states
\beq
  \ket{\alpha}=\sum_{\mu \in \E}
               {\psi_{\alpha}}^{\mu} \dket{\mu}, \quad \quad
               \alpha \in \V.
\label{eq:boundary}
\eeq
Here $\E$ is a set labeling the Ishibashi states allowed
in the model. 
In general, $\E$ does not contain all the elements 
appearing in the set $\I$ of the possible representations.
Rather, $\E$ is a set distinct from $\I$
since the multiplicity of a representation $\mu$ in $\E$ can be greater
than 1, as is seen in the $su(2)$ $D_{\text{even}}$ invariant.

The coefficients ${\psi_{\alpha}}^{\mu}$ 
together with the sets $\E$
and $\V$ should be chosen appropriately
for the mutual consistency of the boundary states.
Consider the annulus amplitude between two 
boundary states $\ket{\alpha}$ and $\ket{\beta}$,
\beq
 Z_{\alpha \beta}=\bra{\beta}\tilde{q}^{H_c}\ket{\alpha}
                 = \sum_{\nu \in \E, 
                         \mu \in \I}
                   {\psi_{\alpha}}^{\nu}
                   \frac{S_{\mu \nu}}
                        {S_{0 \nu}}
                   \overline{{\psi_{\beta}}^{\nu}}\,
                   \chi_{\mu}(\tau), \quad\quad
  \alpha, \beta \in \V .
\eeq
We denote by $n_{\mu \alpha}{}^{\beta}$ the multiplicity
of the representation $\mu \in \I$
in $Z_{\alpha \beta}$,
\beq
  {n_{\mu \alpha}}^{\beta}
     =\sum_{\nu \in \E} 
     {\psi_{\alpha}}^{\nu}
                   \frac{S_{\mu \nu}}
                        {S_{0 \nu}}
                   \overline{{\psi_{\beta}}^{\nu}}
     =\sum_{\nu \in \E} 
     {\psi_{\alpha}}^{\nu}
                   \gamma^{(\mu)}_{\nu}
                   \overline{{\psi_{\beta}}^{\nu}}, 
\label{eq:Cardy}
\eeq
where $\displaystyle \gamma^{(\mu)}_\nu$ is the generalized quantum dimension
\beq
 \gamma^{(\mu)}_{\nu}
             =\frac{S_{\mu \nu}}{S_{0 \nu}}.
\label{eq:qmdim}
\eeq
Clearly, ${n_{\mu \alpha}}^{\beta}$ takes
non-negative integer values for consistent boundary states.
In addition to this, ${n_{0 \alpha}}^{\beta}=\delta_{\alpha \beta}$
since the vacuum is unique.
We call this set of conditions the Cardy condition
and the boundary states satisfying these conditions
the Cardy states \cite{Cardy}.
In terms of the matrix notation 
\beq
  {(\psi)_{\alpha}}^{\mu}={\psi_{\alpha}}^{\mu}, \quad
  {(n_\mu)_{\alpha}}^{\beta}={n_{\mu \alpha}}^{\beta}, \quad
  \gamma^{(\mu)}=\text{diag}(\gamma^{(\mu)}_\nu),
\eeq
the Cardy condition is summarized as follows: 
\bes
   n_{\mu} &= \psi \gamma^{(\mu)} \psi^{\dagger} 
                         \quad    \in \mathrm{Mat}(\abs{\V},\Z_{\ge 0}),\\
   n_{0}   &= \psi \psi^{\dagger} = 1 .
\label{eq:Cardymat}
\ees
Here $\mathrm{Mat}(m,\Z_{\ge 0})$ is the set of 
$m \times m$ matrices with non-negative integer entries.

So far, the number of the independent Cardy states $\abs{\V}$
is not specified.
Hereafter, we assume that
the number of the Cardy states is equal to the
number of the Ishibashi states \cite{Pradisi}
\beq
  \abs{\V} = \abs{\E} \quad\quad \text{(completeness)}.
\label{eq:completeness}
\eeq
If this holds, $\psi$ is a square matrix, and 
the Cardy condition \eqref{eq:Cardymat} means that $\psi$ is unitary.
The situation is quite analogous to the Verlinde 
formula \cite{Verlinde}
\beq
  {\mathcal{N}_{\mu \nu}}^{\rho} 
     = \sum_{\lambda \in \I}
       \frac{S_{\mu \lambda}S_{\nu \lambda}
            \overline{S_{\rho \lambda}}}
            {S_{0\lambda}}
     = \sum_{\lambda \in \I}
            S_{\nu \lambda}\gamma^{(\mu)}_{\lambda}
            \overline{S_{\rho \lambda}},
\label{eq:Verlinde}
\eeq
where $\fusion{\mu}{\nu}{\rho}$
is the fusion coefficient $(\mu)\times (\nu) 
= \sum_{\rho \in \I} \fusion{\mu}{\nu}{\rho}(\rho)$.
In the matrix notation
\beq
  (N_\mu)_\nu{}^\rho = \fusion{\mu}{\nu}{\rho}, 
\label{eq:regularNIM}
\eeq
the above formula can be written as
\bes
     {N}_{\mu} &= S \gamma^{(\mu)} S^{\dagger}
                         \quad    \in \mathrm{Mat}(\abs{\I},\Z_{\ge 0}),\\
     {N}_{0}   &= SS^{\dagger} = 1 .
\label{eq:Verlindemat}
\ees
From the associativity of the fusion algebra, one can show that
$N_\lambda$ satisfies the fusion algebra
\beq
  N_\mu N_\nu = \sum_{\rho \in \I} \fusion{\mu}{\nu}{\rho} N_\rho .
\label{eq:Nfusion}
\eeq
Applying the Verlinde formula \eqref{eq:Verlindemat} to both sides,
one obtains
\beq
  \gamma^{(\mu)} \gamma^{(\nu)} = 
  \sum_{\rho \in \I} \fusion{\mu}{\nu}{\rho} \gamma^{(\rho)} .
\label{eq:qmdimfusion}
\eeq
The generalized quantum dimension
$\{\gamma^{(\mu)}_\lambda |\, \lambda \, \text{fixed}\}$ is therefore
a one-dimensional representation of the fusion algebra.
If we use $\psi$ instead of $S$ in the Verlinde formula,
we obtain $n_\mu$. 
Hence, $n_\mu$, as well as $N_\mu$, satisfies the fusion algebra
\beq
  n_\mu n_\nu = \sum_{\rho \in \I} \fusion{\mu}{\nu}{\rho} n_\rho .
\label{eq:nfusion}
\eeq
The Cardy condition \eqref{eq:Cardymat} together with
the assumption of completeness \eqref{eq:completeness} implies that
$\{n_\lambda\}$ forms a non-negative integer matrix representation
(NIM-rep) of the fusion algebra \cite{BPPZ}.

For each set of the mutually consistent boundary states,
we have a NIM-rep $\{n_\mu | \, \mu \in \I \}$ of the fusion algebra.
The simplest example is the regular NIM-rep
\bes
  & \psi = S   \quad (\E=\V=\I),\\
  & n_{\mu}=N_{\mu},
\label{eq:Atype}
\ees 
which corresponds to the diagonal modular invariant
and exists in any RCFT.
However, the converse is in general not true. 
There are many `unphysical' NIM-reps that do not correspond
to any modular invariant \cite{Gannon}.
The typical example is the tadpole NIM-rep $T_n$ of $su(2)_{2n-1}$
\cite{DZ,BPPZ}.
This fact shows that the Cardy condition is not a sufficient
but a necessary condition for consistency.

\subsection{Action of simple currents}
\label{sec:sconNIM}

We next argue the action of the simple currents on NIM-reps,
which is of fundamental importance in the construction
of boundary states in coset CFTs \cite{Ishikawa,GG}.

A simple current $J$ of a RCFT is a representation whose
fusion with the other representations induces 
a permutation $\mu \mapsto J \mu$ of $\I$
\cite{SY,Intriligator}, 
\beq
  (J) \times (\mu) = (J \mu) .
\eeq
The set of all the simple currents forms an abelian group 
which we denote by $\G$,
\beq
  \G = \{J | \,\,\text{simple currents}\}.
\eeq
$\G$ is a multiplicative group in the fusion algebra.
The group multiplication is defined by the fusion.

A simple current $J$ acts on the modular $S$ matrix as follows,
\beq
  S_{J \lambda\, \mu} = S_{\lambda \mu} b_\mu(J),
\label{eq:sdual}
\eeq
where $b_\mu(J)$ is the generalized quantum dimension of $J$,
\footnote{%
We write $J0$ as $J$ 
since the action of $J$ on $0$ yields $J$ itself.
}
\beq
  b_\mu(J) = \gamma^{(J)}_\mu = \frac{S_{J \mu}}{S_{0 \mu}} .
\label{eq:bandqmdim}
\eeq
In the matrix notation, the above relation can be
written as
\beq
  N_J S = S \, b(J) ,
\label{eq:sdualmat} 
\eeq
where
$(N_J)_\mu{}^\nu = \fusion{J}{\mu}{\nu} = \delta_{J\!\mu\,\nu}$
is the regular NIM-rep \eqref{eq:regularNIM}
and $b(J) = \mathrm{diag}(b_\mu(J))$.
The transformation property \eqref{eq:sdualmat} readily follows
from the Verlinde formula
\beq
  N_J = S \gamma^{(J)} S^\dagger = S \,b(J) S^\dagger .
\eeq
Since $\displaystyle (\lambda) \mapsto \gamma^{(\lambda)}_\mu$
is a one-dimensional representation of the fusion algebra,
$J \mapsto b_\mu(J)$ is a one-dimensional representation of $\G$,
\beq
  b_\mu(J J') = b_\mu(J) b_\mu(J') , \quad J, J' \in \G .
\eeq
Therefore $b_\mu(J)$ is of the form
\beq
  b_\mu(J) = e^{2\pi i Q_\mu(J)} , \quad Q_\mu(J) \in \Q ,
  \label{eq:bQ}
\eeq
since any one-dimensional representation of a finite group takes
values in roots of unity.
The phase $Q_\mu(J)$ is called the monodromy charge.

The similarity between $S_{\mu \nu}$ and $\psi_{\alpha}{}^{\mu}$
\eqref{eq:Cardymat}\eqref{eq:Verlindemat} suggests that
the simple currents act also on a diagonalization matrix $\psi$.
For $\psi_\alpha{}^\mu$, however, 
we can consider two types of actions of the simple currents,
since $\alpha$ and $\mu$ take values in the different sets $\V$ and $\E$,
respectively.

First, we consider the action of the simple currents on
the label $\alpha \in \V$ of the Cardy states.
In order to see this, 
we rewrite the Cardy condition \eqref{eq:Cardymat} in the form
\beq
 n_\nu \psi = \psi \gamma^{(\nu)}, \quad \nu \in \I.
\eeq
This is possible since $\psi$ is a unitary matrix. 
Setting $\nu = J \in \G \subset \I$, we obtain
\beq
 n_J \psi = \psi \gamma^{(J)} = \psi b(J) , 
\eeq
where we used the relation \eqref{eq:bandqmdim}.
In the component form, this equation can be written as
\beq
  \sum_{\beta \in \V} (n_J)_\alpha{}^{\beta} \psi_{\beta}{}^\mu =
  \psi_\alpha{}^\mu b_\mu(J) , \quad \mu \in \E.
  \label{eq:JV1}
\eeq
We can regard this as a relation among the (row) vectors
$\{(\psi_\alpha{}^\mu)_{\mu \in \E} | \alpha \in \V \}$.
Namely, the vector
$\psi_\alpha b(J) = (\psi_\alpha{}^\mu \, b_\mu(J))_{\mu \in \E}$
is expressed as a linear combination of
the vectors $\{\psi_\beta | \beta \in \V \}$ with
the non-negative integer coefficients
$(n_J)_\alpha{}^{\beta} \in \Z_{\ge 0}$. 
One can estimate the number of vectors contributing to the sum
in eq.\eqref{eq:JV1} by calculating the length of
the vector $\psi_\alpha b(J)$,
\beq
  \norm{\psi_\alpha b(J)}^2 =
  \sum_{\mu \in \E} \psi_\alpha{}^\mu \, b_\mu(J) 
  \overline{\psi_\alpha{}^\mu \, b_\mu(J)} =
  \sum_{\mu \in \E} \psi_\alpha{}^\mu \overline{\psi_\alpha{}^\mu} = 1,
  \label{eq:normJalpha}
\eeq
where we used \eqref{eq:bQ}.
This means that there is exactly one vector in the sum \eqref{eq:JV1}
since the vectors $\{\psi_\beta | \beta \in \V\}$ form
an orthonormal set with respect to the inner product in
\eqref{eq:normJalpha} and the coefficients $(n_J)_\alpha{}^\beta$ 
take non-negative integer values.
In other words, the vector $\psi_\alpha b(J)$ coincides with
one of the vectors $\{\psi_\beta | \beta \in \V \}$,
which we denote by $\psi_{J \alpha} (J \alpha \in \V)$.
With this notation, eq.\eqref{eq:JV1} can be written as follows:
\beq
  \psi_{J \alpha}{}^\mu \equiv
  \sum_{\beta \in \V} (n_J)_\alpha{}^{\beta} \psi_{\beta}{}^\mu =
   \psi_\alpha{}^\mu \, b_\mu(J), \quad
  \alpha, J \alpha \in \V , \, J \in \G . 
\label{eq:JV}
\eeq
The map $\alpha \mapsto J\alpha$ is one-to-one and onto, i.e.,
$n_J$ is a permutation matrix.
Actually, there exists a positive integer $m$ such that
$(n_J)^m = 1$ since $\{n_J | J \in \G\}$ is a NIM-rep of
the simple current group $\G$, which is in general 
a product of cyclic groups.
This means that $J^m \alpha = \alpha, \forall \alpha \in \V$.
We can therefore conclude that
$J \in \G$ acts on $\V$ as a permutation $\alpha \mapsto J\alpha$.

There may be some element $J_0 \in \G$ such that $J_0 \alpha = \alpha$ for
all $\alpha \in \V$.
From \eqref{eq:JV}, this condition is equivalent to
$b_\mu(J_0) = 1$  for all $\mu \in \E$.
This is possible because the set $\E$ is distinct from $\I$.
These elements $J_0 \in \G$ form a subgroup $\stab{\V}$
of $\G$ which we call
the stabilizer of $\V$
\beq
  \stab{\V}
     =\{ J_0 \in \G \,|\, J_0 \alpha = \alpha ,\, 
                         \forall \alpha \in \V \}
     =\{ J_0 \in \G \,|\, b_{\mu}(J_0)=1 ,\, 
                         \forall \mu \in \E \}. 
  \label{eq:stabilizer}
\eeq 
Since the stabilizer acts on $\V$ trivially, it is natural to
consider the quotient of $\G$ by $\stab{\V}$.
We denote by $\auto{\V}$ the quotient group
and call it the group of automorphisms of $\V$,
\beq
  \auto{\V}= \G /\stab{\V}.
  \label{eq:autoV}
\eeq

We next turn to the action of the simple currents 
on the label $\mu \in \E$ of the Ishibashi states.
Namely, we consider the transformation property of $\psi_\alpha{}^\mu$
under $\mu \mapsto J \mu$. 
For this transformation to make sense, we have to take $J \in \G$
such that $J$ leaves the set $\E$ invariant.
The elements $J \in \G$ that leave $\E$ invariant, i.e.,
can be restricted to $\E$, form a subgroup of $\G$.
We denote it by $\auto{\E}$ and call it the group of automorphisms of $\E$,
\beq
  \auto{\E} = \{ J\in \G \,|\, J: \, \E \rightarrow \E\}.
  \label{eq:autoE}
\eeq
In general, $\auto{\E} \neq \G$ since
there may be some representation $\mu \in \I$
missing in $\E$.

With this definition of $\auto{\E}$, 
we propose the following transformation property of
$\psi$ under $\mu \mapsto J\mu$, 
\beq
  \psi_{\alpha}{}^{J\mu} = \tilde{b}_\alpha(J) \, \psi_\alpha{}^\mu ,
  \quad \mu, J\mu  \in \E,\, J \in \auto{\E} .
\label{eq:JE}
\eeq
Here, $\tilde{b}_\alpha(J)$ is the counterpart of $b_\mu(J)$
in eq.\eqref{eq:JV} and determined by $\alpha \in \V$ and $J \in \auto{\E}$.
Unlike eq.\eqref{eq:JV} on $\V$,
the present authors have no rigorous proof of 
the above equation \eqref{eq:JE} on $\E$.
Hence, we \textit{assume} in this paper 
that the transformation property \eqref{eq:JE} holds.
This is not so restrictive. 
Actually, all the examples treated in this paper satisfy \eqref{eq:JE}
with an appropriate choice of $\tilde{b}_\alpha(J)$. 

To summarize, the diagonalization matrix $\psi_\alpha{}^\mu$
transforms under the action of the simple currents as follows,
\begin{subequations}
\label{eq:sconNIM}
\bea
  \psi_{J \alpha}{}^\mu &= \psi_\alpha{}^\mu \, b_\mu(J), &
  n_J \psi &= \psi \, b(J) , &
  J &\in \auto{\V} , 
  \label{eq:sconbrane} \\
  \psi_{\alpha}{}^{J\mu} &= \tilde{b}_\alpha(J) \, \psi_\alpha{}^\mu, &
  \psi N_J^T &= \tilde{b}(J) \psi , &
  J &\in \auto{\E}.
  \label{eq:sconspec}
\eea
\end{subequations}
Here we use the same symbol $N_J$ for its restriction to $\E$.

For a NIM-rep $n_{\mu}$ of the fusion algebra,
there corresponds a graph whose vertices are
labelled by the set $\V$ \cite{DZ,BPPZ}.
We can identify the boundary states with the vertices of
the graph. Then the automorphism $\auto{\V}$
is naturally interpreted as the automorphism 
of the graph, while $\tilde{b}_{\alpha}$ represents
a coloring of the graph.

\bigskip
\subsection{Construction of NIM-reps}
\label{sec:NIMconstruction}

In this subsection, we present two methods to yield a new NIM-rep
from a given one, which are used to construct
non-trivial NIM-reps in coset theories.
One is the use of the simple currents
\cite{FHSSW,GG},  which is nothing but
the NIM-rep version of the orbifold construction.
The other is based on the conformal embedding\footnote{
The NIM-reps associated with conformal embeddings has
been studied also from the operator-algebraic point of
view~\cite{Xu,BE,BEK,FuchsSchweigert}.}.

\subsubsection{Simple currents}
\label{sec:NIMsc}

The transformation property \eqref{eq:sconbrane} of 
NIM-reps enables us to take the orbifold of a NIM-rep
$n_\mu = \psi \gamma^{(\mu)} \psi^\dagger$
by the simple currents $J \in \auto{\V}$.
Suppose that we have a non-trivial subgroup
$H \subset \auto{\V}$.
By the action of $H$, the set $\V$ splits into several orbits,
which we denote by $\tilde{\V} = \{[\alpha], \alpha \in \V\}$.
For simplicity, we assume that the action of $H$ has no fixed points
and that all the orbits have the same length $\abs{H}$.
Then, we can construct a new NIM-rep $\tilde{\psi}$ 
by orbifolding $\psi$,
\beq
 \tilde{\psi}_{[\alpha]}{}^\mu = 
 \frac{1}{\sqrt{\abs{H}}} \sum_{J \in H} \psi_{J\alpha}{}^\mu = 
 \frac{1}{\sqrt{\abs{H}}} \psi_{\alpha}{}^\mu \sum_{J \in H} b_\mu(J) .
\eeq
The factor $\sum_{J \in H} b_\mu(J)$ vanishes unless
$b_\mu(J) = 1$ for all $J \in H$, since $b_\mu(J)$ is one-dimensional
representation of the permutation group $H$.
In other words, $\sum_{J \in H} b_\mu(J)$ projects $\E$ to
the set
\beq
  \tilde{\E} = \{\mu \in \E \, |\, b_\mu(J)=1, \, \forall J \in H\},
  \label{eq:scE}
\eeq
which is the set of the representations $\mu$ with vanishing monodromy
charge with respect to $H \subset \G$.
On the set $\tilde{\E}$, $\tilde{\psi}$ takes the simple form
\cite{FHSSW,GG}
\beq
 \tilde{\psi}_{[\alpha]}{}^\mu = \sqrt{\abs{H}} \psi_\alpha{}^\mu , \quad
 [\alpha] \in \tilde{\V} , \, \mu \in \tilde{\E} .
 \label{eq:scNIM}
\eeq

We shall show that $\tilde{\psi}$ together with
$\tilde{\V}$ and $\tilde{\E}$ defines a new NIM-rep
$\tilde{n}_\mu = \tilde{\psi} \gamma^{(\mu)} \tilde{\psi}^\dagger$. 
For $\tilde{n}_\mu$ to be a NIM-rep, it is necessary 
that $\tilde{\psi}$ is a square matrix.
This follows from the assumption that
the action of $H$ on $\V$ has no fixed points. 
Since $H$ is a finite abelian group,
$H$ can be written as a product of cyclic groups,
\beq
  H = \Z_{m_1} \times \Z_{m_2} \times \cdots .
\eeq
Each factor $\Z_{m_j}$ acts as a cyclic permutation of $m_j$ elements since
the action has no fixed points. 
Hence $b(J_j) = \psi^\dagger n_{J_j} \psi$ 
for the generator $J_j$ of $\Z_{m_j}$ gives all the possible
eigenvalues of the cyclic permutation $\Z_{m_j}$, 
which takes values in roots of unity.
More precisely, the set $\{b_\mu(J_j), \mu \in \E \}$ for the generator
of $\Z_{m_j}$ is $\abs{\E}/m_j$ copies of
$\{1, e^{2\pi i/m_j} , \ldots, e^{2\pi i(m_j - 1)/m_j}\}$
(one copy for one orbit in $\V$ of $\Z_{m_j}$).
Since $b_\mu(J_j) = 1$ is the necessary and
sufficient condition for $b_\mu(J) = 1 (\forall J \in \Z_{m_j})$,
we have an identity
$\abs{\{\mu \in \E | b_\mu(J) = 1, \forall J \in \Z_{m_j}\}} = \abs{\E}/m_j$.
Repeating this argument for each factor $\Z_{m_j}$ of $H$,
we obtain
\beq
  \abs{\tilde{\E}} = \abs{\E}/(m_1 m_2 \cdots) = \abs{\E}/\abs{H}
  = \abs{\V}/\abs{H} = \abs{\tilde{\V}} ,
\eeq
which shows that $\tilde{\psi}$ is a square matrix.
Then it follows immediately that
$\tilde{n}_\mu = \tilde{\psi} \gamma^{(\mu)} \tilde{\psi}^\dagger$
yields a NIM-rep defined in $\tilde{\V}$,
\bes
  \tilde{n}_{\mu [\alpha]}{}^{[\beta]}
  = \sum_{\nu \in \tilde{\E}}
  \tilde{\psi}_{[\alpha]}{}^\nu \gamma^{(\mu)}_\nu
  \overline{\tilde{\psi}_{[\beta]}{}^\nu}
  &= \abs{H} \sum_{\nu \in \tilde{\E}}
  \psi_{\alpha}{}^\nu \gamma^{(\mu)}_\nu
  \overline{\psi_{\beta}{}^\nu} \\
  &= \abs{H} \sum_{\nu \in \E}
  \frac{1}{\abs{H}} \sum_{J \in H} b_\nu(J)
  \psi_{\alpha}{}^\nu \gamma^{(\mu)}_\nu
  \overline{\psi_{\beta}{}^\nu} \\
  &= \sum_{J \in H} 
  \sum_{\nu \in \E}
  \psi_{J\alpha}{}^\nu \gamma^{(\mu)}_\nu
  \overline{\psi_{\beta}{}^\nu} \\
  &=  \sum_{J \in H} n_{\mu J\alpha}{}^\beta .
\ees
Here we used the action \eqref{eq:sconbrane} of the simple currents
on $\psi$.
From this equation, $\tilde{n}_0$ is the identity matrix
in $\tilde{\V}$ and $\tilde{\psi}$ is unitary.
Since the entries of $\tilde{n}_\mu$ are manifestly non-negative
integers, $\{\tilde{n}_\mu\}$ forms a NIM-rep of the fusion algebra in
$\tilde{\V}$. 
One can consider this NIM-rep as corresponding to 
the orbifold of the charge conjugation modular invariant,
since the spectrum $\tilde{\E}$ of the Ishibashi states is the same
as that of the orbifold. 

Of course, the above construction should be modified
if there are fixed points in the action of $H$ on 
the set $\V$ of the Cardy states.
This is the familiar situation in the orbifold construction,
and we need extra Ishibashi states to resolve the fixed points in $\V$.

\subsubsection{Conformal embedding}
\label{sec:embedding}

Consider an embedding ${\tilde{g}} \subset {g}$
of the affine Lie algebra $\tilde{g}$ into $g$.
An embedding ${\tilde{g}} \subset {g}$ is
called conformal if the conformal invariance of the theory
is preserved.
For a conformal embedding, 
the energy-momentum tensors of the corresponding CFTs
(the $G$ and $\widetilde{G}$ WZW models for the simple algebras) 
are the same.
In particular, the central charges of two theories match
\beq
   c({\tilde{g}}) = c({g}) .
\eeq
This condition is sufficient for an embedding to be conformal.
All the conformal embeddings has been classified in \cite{SW,BB}.

Let $\tilde{g} \subset g$ be a conformal embedding.
Since the $g$-theory is isomorphic to the $\tilde{g}$-theory as a CFT,
the conformal embedding ${\tilde{g}} \subset {g}$
enables us to regard the boundary states of the $g$-theory 
as those of the $\tilde{g}$-theory.
In order to see this, we consider the branching rule of the representations,
\beq
  (\mu) \mapsto \dirsum_{\tilde{\mu} \in \tilde{\E}_{\mu}} (\tilde{\mu}) ,
  \quad \mu \in \I, 
  \quad \tilde{\mu} \in \tilde{\I}.
\label{eq:branching}
\eeq
Here we denote by $\tilde{\E}_\mu$ the set of the representations
in the $\tilde{g}$-theory that $\mu \in \I$ branches to.
$\I$ and $\tilde{\I}$ are the set of the integrable representations
in the $g$ and $\tilde{g}$ theories, respectively.
According to the branching rule \eqref{eq:branching}, 
the Ishibashi states of the $g$-theory can be expressed
in terms of those of the $\tilde{g}$-theory,
\beq
   \dket{\mu}
         = \sum_{\tilde{\mu} \in \tilde{\E}_{\mu}} 
                  \sqrt{\frac{\tilde{S}_{0 \tilde{\mu}}}
                       {S_{0 \mu}}} \,
           \dket{\tilde{\mu}},
   \label{eq:Ishibashibranching}
\eeq
where $S$ and $\tilde{S}$ are the modular $S$-matrix of the $g$ and
$\tilde{g}$-theories, respectively.
The coefficients of $\dket{\tilde{\mu}}$ arise
due to the difference of the normalization \eqref{eq:Ishibashinorm} of
the Ishibashi states in the $g$ and $\tilde{g}$-theories.
To be precise, we can multiply the coefficients by some phases
without changing the normalization. 
We shall see, however, that these phases does not influence
the resulting NIM-rep (see eq.\eqref{eq:phaseambiguity}).

Let 
$\ket{\alpha} = \sum_{\mu \in \E} \psi_\alpha{}^\mu \dket{\mu}\,
(\alpha \in \V)$
be the Cardy states of the $g$-theory.
Using the above expression for the Ishibashi states,
one can rewrite the Cardy state $\ket{\alpha}$ of the $g$-theory
as a linear combination of the Ishibashi states of the $\tilde{g}$-theory
as follows,
\bes
  \ket{\alpha} = \sum_{\mu \in \E} \psi_\alpha{}^\mu \dket{\mu} 
  &= \sum_{\mu \in \E} \psi_\alpha{}^\mu 
  \sum_{\tilde{\mu} \in \tilde{\E}_{\mu}} 
                  \sqrt{\frac{\tilde{S}_{0 \tilde{\mu}}}{S_{0 \mu}}} \,
           \dket{\tilde{\mu}} \\
  &=  \sum_{\tilde{\mu} \in \tilde{\E}} \tilde{\psi}_\alpha{}^{\tilde{\mu}}
     \dket{\tilde{\mu}} .
\label{eq:Etype}
\ees
Here we introduced the set
\beq
 \tilde{\E} = \coprod_{\mu \in \E}
             \tilde{\E}_{\mu} ,
\label{eq:embedspec}
\eeq
and denoted by $\tilde{\psi}_\alpha{}^{\tilde{\mu}}$ the coefficients
of the Ishibashi states
\beq
  \tilde{\psi}_\alpha{}^{\tilde{\mu}}
  =  \sqrt{\frac{\tilde{S}_{0 \tilde{\mu}}}{S_{0 \mu}}}
     \psi_\alpha{}^\mu \quad 
  (\alpha \in \V, \quad \tilde{\mu} \in \tilde{\E}_\mu).
  \label{eq:embedpsi}
\eeq
Hence, the Cardy states $\ket{\alpha} \, (\alpha \in \V)$ of the $g$-theory
can be regarded as boundary states of the $\tilde{g}$-theory
with the coefficients
$\tilde{\psi}_\alpha{}^{\tilde{\mu}}$.

Note that a representation $\tilde{\mu} \in \tilde{\I}$
may appear more than once in $\tilde{\E}$, 
since it is possible that two different representations
$\mu_1, \mu_2 \in \I$ contain the same representation
$\tilde{\mu} \in \tilde{\I}$.
When this is the case, we have to treat these $\tilde{\mu}$'s as elements
orthogonal with each other in $\tilde{\mathcal{E}}$,
since they constitute distinct representations of the $g$-theory. 
\footnote{%
An example is $su(2)_4 \subset su(3)_1$.}

The boundary states of the $\tilde{g}$-theory obtained in this way
do not satisfy the completeness condition \eqref{eq:completeness} since
$\abs{\V} = \abs{\E} < \abs{\tilde{\E}}$.
This is because a representation $\mu \in \E$ of the $g$-theory in general
branches to more than one representations in the $\tilde{g}$-theory.
Therefore we need additional states other than those in $\V$ 
to satisfy the completeness condition.
In other words, there are some states missing in $\V$ 
for $\tilde{\psi}$ to be a NIM-rep of the $\tilde{g}$-theory.
From the point of view of the $g$-theory,
the missing states break the boundary condition of the $g$-theory
that $\ket{\alpha} \, (\alpha \in \V)$ obeys.

In many cases, we can generate the missing states from those in $\V$
to construct a NIM-rep of the $\tilde{g}$-theory.
Suppose that we have a NIM-rep
$\tilde{\psi}_{\tilde{\alpha}}{}^{\tilde{\mu}}
\,(\tilde{\alpha} \in \tilde{\V}, \, \tilde{\mu} \in \tilde{\E})$
of the $\tilde{g}$-theory such that
the matrix \eqref{eq:embedpsi} constitutes a part of $\tilde{\psi}$,
i.e., $\V \subset \tilde{\V}$.
The Cardy condition \eqref{eq:Cardymat} for the $\tilde{g}$-theory
can be written as
\beq
  \tilde{\psi}_{\tilde{\alpha}}{}^{\tilde{\nu}}
  \gamma^{(\tilde{\mu})}_{\tilde{\nu}} =
  \sum_{\tilde{\beta} \in \tilde{\V}}
  \tilde{n}_{\tilde{\mu} \tilde{\alpha}}{}^{\tilde{\beta}}
  \tilde{\psi}_{\tilde{\beta}}{}^{\tilde{\nu}}, 
 \quad
  \tilde{\mu} \in \tilde{\I}.
\label{eq:boundarygenerate}
\eeq
Since the generalized quantum dimension $\gamma^{(\tilde{\mu})}$ is
the representation of the fusion algebra,
one can regard this equation as describing the `fusion'
of the Cardy state $\ket{\tilde{\alpha}}$
with $\tilde{\mu} \in \tilde{\I}$.
The right-hand side of this equation gives a linear combination
of the Cardy states with non-negative integer coefficients,
which we denote by $\tilde{\mu} \tilde{\alpha}$.
The fusion of the Cardy state with $\tilde{\mu} \in \tilde{\I}$ therefore
yields another Cardy states in $\tilde{\V}$.
This is a tool appropriate for our purpose
of generating the full Cardy states 
from a part of them, since the generalized quantum dimension
is determined solely by the modular $S$-matrix and
independent of $\tilde{\psi}$.

The construction of the missing states proceeds as follows.
We first take one of the boundary states $\alpha \in \V$ which
originates from those in the $g$-theory. 
Then, we perform the fusion of $\alpha$ with one of
the generators $\tilde{\mu}_1 \in \tilde{\I}$ of the fusion algebra
(For $\tilde{g} = su(2)_k$, $\tilde{\mu}_1$ is the fundamental representation).
According to eq.\eqref{eq:boundarygenerate}, the fusion with $\tilde{\mu}_1$
yields a linear combination of the boundary states
$\tilde{\mu}_1 \alpha$.
In general, $\tilde{\mu}_1 \alpha$ is composed of several Cardy states.
The number of the Cardy states contained in $\tilde{\mu}_1 \alpha$
can be read off from the length of $\tilde{\mu}_1 \alpha$ as a vector
in the space of the boundary states,
in the same way as eq.\eqref{eq:normJalpha}.
Let $\norm{\tilde{\mu}_1 \alpha}$ be the length of $\tilde{\mu}_1 \alpha$
\beq
  \norm{\tilde{\mu}_1 \alpha}^2 = 
  \sum_{\tilde{\nu} \in \tilde{\E}}
  \left|
  \tilde{\psi}_\alpha{}^{\tilde{\nu}}
  \gamma^{(\tilde{\mu}_1)}_{\tilde{\nu}} \right|^2 .
\eeq
If $\norm{\tilde{\mu}_1 \alpha}^2 =1$, $\tilde{\mu}_1 \alpha$
has to coincide with one of the Cardy states, since the coefficients
$\tilde{n}$ in eq.\eqref{eq:boundarygenerate} are non-negative integer
and the Cardy states form an orthonormal basis in the space of
the boundary states.
Hence, for $\norm{\tilde{\mu}_1 \alpha}^2 =1$,
there are two possibilities: 
$\tilde{\mu}_1 \alpha \in \V$, or
$\tilde{\mu}_1 \alpha \notin \V$.
If $\tilde{\mu}_1 \alpha \notin \V$, 
$\tilde{\mu}_1 \alpha$ is a new state in $\tilde{\V}$. 
If $\norm{\tilde{\mu}_1 \alpha}^2 \ge 2$, $\tilde{\mu}_1 \alpha$
contains more than one states, which may include
a new state in $\tilde{\V}$. 
We can subtract the contribution of the known states from
$\tilde{\mu}_1 \alpha$, which is given by calculating the inner product of
$\tilde{\mu}_1 \alpha$ with the known states. 
If the remaining part is a vector of unit length, we
obtain a new state in $\tilde{\V}$.

One can repeat this procedure for all the known states in $\tilde{\V}$ and 
all the generators $\tilde{\mu} \in \tilde{\I}$
until no new states are generated. 
Since $\abs{\tilde{\V}}$ is finite, this procedure terminates
in the finite number of steps.
In this way, we can generate a set of the Cardy states
of the $\tilde{g}$-theory starting from those of the $g$-theory.
It is not clear for the present authors whether
our procedure always yields a complete set of the Cardy states.
However, if the resulting states form a complete set, i.e.,
$\tilde{\V} = \tilde{\E}$, 
it is manifest that they form a NIM-rep of the $\tilde{g}$-theory, 
since our procedure is based on the Cardy 
condition \eqref{eq:boundarygenerate}.
(All the examples in this paper satisfy the completeness.)
One can consider the resulting NIM-rep as corresponding to the exceptional
modular invariant originated from the same conformal embedding
$\tilde{g} \subset g$, since the spectrum $\tilde{\E}$ of
the Ishibashi states coincides with that of the exceptional invariant.

We comment on the ambiguity in the formula 
\eqref{eq:Ishibashibranching} for the branching of the Ishibashi states.
This ambiguity causes some $\tilde{\mu}$-dependent phases
in the resulting diagonalization matrix $\tilde{\psi}$
(see eq.\eqref{eq:embedpsi}).
These phases, however, have no influence on the associated NIM-rep
$\tilde{n} = \tilde{\psi} \gamma \tilde{\psi}^\dagger$.
Actually, if we have another diagonalization matrix
\beq
  \tilde{\psi}' = \tilde{\psi} D, \quad
  D = \mathrm{diag}(e^{i a_{\tilde{\mu}}})_{\tilde{\mu} \in \tilde{\E}}\quad
  (a_{\tilde{\mu}} \in \R) , 
\eeq
the corresponding NIM-rep $\tilde{n}'$ coincides with 
$\tilde{n} = \tilde{\psi} \gamma \tilde{\psi}^\dagger$, 
\beq
  \tilde{n}' = \tilde{\psi}' \gamma (\tilde{\psi}')^\dagger
  = \tilde{\psi} D \gamma D^\dagger \tilde{\psi}^\dagger 
  = \tilde{\psi} \gamma \tilde{\psi}^\dagger = \tilde{n} .
\label{eq:phaseambiguity}
\eeq
Here we used that $\gamma$ commutes with the diagonal matrix $D$.

As an illustration of our procedure, we describe in Appendix A
the construction of the $E_6$ NIM-rep \cite{DZ,BPPZ}
of the $SU(2)$ WZW model, which originates from the conformal embedding
$su(2)_{10} \subset sp(4)_1$.


\section{Boundary states in tensor product theories}
\label{sec:tensor}

In this section, we consider NIM-reps in
tensor product theories $G \times H$.
\footnote{%
In this paper, we restrict ourselves to the tensor products
of the WZW models, though 
the following construction is applicable to any rational
CFTs.}
As we shall show in the next section, 
we can construct 
a large class of NIM-reps in coset theories starting from
those in tensor product theories.
In this sense, NIM-reps in tensor product theories
deserves a detailed study, although it is important in its own right.

\subsection{Preliminaries}

We first summarize the general property of
tensor product theories.
The $G\times H$ theory at level $(k_G,k_H)$ is 
the tensor product of the $G$ WZW model at level
$k_G$ and the $H$ WZW model at level $k_H$.
We therefore begin with the WZW models.

The set of primary fields in the $G$ WZW model
is given by the set $P_+^{k_G}(g)$
of all the integrable representations 
of the affine Lie algebra $g$ at level $k_G$.
We denote by $S^{G}_{\mu\nu}$
the modular $S$-matrix of the $G$ WZW model.
The simple current group of the $G$ WZW model is
the normal abelian subgroup $\OO{g}$ of 
the outer automorphism group of the affine Lie algebra $g$
\cite{Bernard}.
The action of the simple current $J\in \OO{g}$
on a representation $\mu$ yields 
\beq
   S^{G}_{J\!\mu \,\nu}
       = S^{G}_{\mu \nu}b^{G}_{\nu}(J), \quad J \in \OO{g}, 
\label{eq:Gsdual}
\eeq
with 
\beq
  b^{G}_{\mu}(J)   =    e^{-2\pi i (J\Lambda_0,\mu)}.
\label{eq:monodromy}
\eeq
Here $\Lambda_0$ is the $0$-th fundamental weight of $g$.
In the $G$ WZW model, $b^G$ is an isomorphism
from $\OO{g}$ to the center $Z(G)$ of the group $G$
\beq
\left.
\begin{array}{cccc} 
  b^{G} :& \OO{g} & \xrightarrow{\,\,\cong\,\,}
                          & Z(G)    \vspace{0.3cm}\\
         &     J          &     \longmapsto          
                      & b^G(J)= e^{-2\pi i J\Lambda_0}.
\end{array}
\right.
\eeq
Therefore, eq.\eqref{eq:Gsdual} is interpreted as
intertwining $\OO{g}$ with $Z(G)$.
We adopt the similar notations for the $H$ WZW model. 

We turn to the $G\times H$ theories.
The set $\I$ of the primary fields of the $G \times H$ theory
is given by the tensor product
of the spectrum of the $G$ and $H$ theories,
\beq
  \I = \{(\mu,\nu)\,|\,     \mu  \in P_+^{k_G}(g),\,
                            \nu \in P_+^{k_H}(h)  \}.
\label{eq:tensorspec}
\eeq
Consequently, the $S$ matrix $S_{(\mu, \nu)(\mu', \nu')}$ of the 
$G \times H$ theory reads
\beq
 S_{(\mu,\nu)(\mu',\nu')} = S^G_{\mu \mu'}S^H_{\nu \nu'}.
 \label{eq:tensorS}
\eeq  
The fusion algebra of the $G \times H$ theory is determined via
the Verlinde formula \eqref{eq:Verlinde} and 
is given by the direct product of the fusion algebras
of the $G$ and $H$ theories.
Hence, the simple current group $\G$ of the $G\times H$ theory
is also given by the direct product of $\OO{g}$ and
$\OO{h}$,
\beq
  \G  = \OO{g} \times \OO{h} = 
  \{ (J,J') \,|\, J \in \OO{g},\,
                  J' \in \OO{h}\},   \label{eq:tensorsc}
\eeq
which acts on $\I$ as
\beq
   (J,J')(\mu,\nu) =  (J\mu,J'\nu). 
\eeq 
The action on the S matrix takes the form
\beq
   S_{(J\mu,J'\nu)(\mu',\nu')}
       = S_{(\mu,\nu)(\mu',\nu')}b^{G}_{\mu'}(J)
                                 b^{H}_{\nu'}(J') , \quad
   (J,J') \in \G . 
\eeq

\subsection{Construction of NIM-reps in tensor product theories}

We can construct a NIM-rep of the $G \times H$ theory 
from those of the $G$ and $H$ theories.
This is because
the fusion algebra of the $G \times H$ theory
is given by the direct product of the fusion algebras of
the $G$ and $H$ theories.
For example,
the regular NIM-rep of the $G \times H$ theory takes the form
\beq
  (N_{(\mu, \nu)})_{(\alpha, \beta)}{}^{(\alpha', \beta')}
   = (N^G_\mu)_\alpha{}^{\alpha'} (N^H_\nu)_\beta{}^{\beta'}, \quad
  (\mu, \nu), (\alpha, \beta) , (\alpha',\beta') \in \I ,
\label{eq:tensorAtype}
\eeq
which readily follows from  the $S$ matrix \eqref{eq:tensorS}.
One can see that 
the regular NIM-rep of the $G \times H$ theory is factorized 
into the product of those for the $G$ and $H$ theories. 
We can extend this structure to more general NIM-reps of
tensor product theories. 
Suppose that we have NIM-reps $n^G$ and $n^H$ 
(see eq.\eqref{eq:Cardy}) of the $G$ and $H$ theories,
respectively,
\bes
  (n^G_\mu)_\alpha{}^{\alpha'} &= 
  \sum_{\mu' \in \E^G} 
  \psi^G_{\alpha}{}^{\mu'} \gamma^{G(\mu)}_{\mu'}
  \overline{\psi^{G}_{\alpha'}{}^{\mu'}} , \quad
  \alpha, \alpha' \in \V^G , \\
  (n^H_\nu)_\beta{}^{\beta'} &= 
  \sum_{\nu' \in \E^H} 
  \psi^H_{\beta}{}^{\nu'} \gamma^{H(\nu)}_{\nu'}
  \overline{\psi^{H}_{\beta'}{}^{\nu'}} , \quad
  \beta, \beta' \in \V^H . \\
\ees
Clearly, the tensor product
\beq
  (n_{(\mu, \nu)})_{(\alpha, \beta)}{}^{(\alpha', \beta')}
  = (n^G_\mu \otimes n^H_\nu)_{(\alpha, \beta)}{}^{(\alpha', \beta')}
  = (n^G_\mu)_\alpha{}^{\alpha'} (n^H_\nu)_\beta{}^{\beta'}
  \quad ( (\mu, \nu) \in \I )
\eeq
of these two matrices
yields a NIM-rep of the $G \times H$ theory.
We denote this NIM-rep by $n^G \otimes n^H$.
The labels of the Cardy states and the Ishibashi
states for the NIM-rep $n^G \otimes n^H$ take the form 
\bes
  \V &= \V^G \otimes \V^H =
  \{(\alpha, \beta) | \alpha \in \V^G, \, \beta \in \V^H \} , \\
  \E &= \E^G \otimes \E^H =
  \{(\mu, \nu) | \mu \in \E^G, \, \nu \in \E^H \} . \\
\ees
The diagonalization matrix of the NIM-rep $n^G \otimes n^H$ is
written as
the tensor product of $\psi^G$ and $\psi^H$,
\beq
  \psi_{(\alpha, \beta)}{}^{(\mu, \nu)} =
  (\psi^G \otimes \psi^H)_{(\alpha, \beta)}{}^{(\mu, \nu)} =
  \psi^{G}_{\,\alpha}{}^\mu \psi^{H}_{\,\beta}{}^{\nu} , \quad
  (\alpha, \beta) \in \V , \quad (\mu, \nu) \in \E .
\eeq
The regular NIM-rep \eqref{eq:tensorAtype} is obtained
by setting $\psi^G = S^G, \psi^H = S^H$. 

It should be noted that, for the tensor product
$n^G \otimes n^H$,  the labels of the Cardy states
has the factorized form $(\alpha, \beta)$.
Of course, this is a structure specific to the tensor product NIM-rep
$n^G \otimes n^H$.
A generic NIM-rep of the $G \times H$ theory takes the form
\beq
   (n_{(\mu, \nu)})_\alpha{}^{\beta}
          = \sum_{(\mu', \nu') \in \E}
            \psi_{\alpha}{}^{(\mu',\nu')}
            \frac{S_{(\mu,\nu)(\mu', \nu')}}{S_{(0,0)(\mu', \nu')}}
            \overline{\psi_{\beta}{}^{(\mu',\nu')}} ,
   \quad \alpha, \beta \in \V .
\label{eq:tensorNIMrep}
\eeq
The action \eqref{eq:sconNIM} of the simple currents \eqref{eq:tensorsc}
on the diagonalization matrix $\psi$ reads
\begin{subequations}
\label{eq:scontensor}
\bea
     {\psi_{(J,J')\alpha}}^{(\mu,\nu)}
    & =  {\psi_{\alpha}}^{(\mu,\nu)}
           b^G_{\mu}(J)b^H_{\nu}(J') ,&& (J,J') \in \auto{\V}, 
 \label{eq:scontensorbrane}                                   \\
  {\psi_{\alpha}}^{(J\mu,J'\nu)}
     & = \tilde{b}_{\alpha}(J,J') 
         {\psi_{\alpha}}^{(\mu,\nu)} ,&& (J,J') \in \auto{\E},
\label{eq:scontensorspec}
\eea
\end{subequations}
where $(J,J')\alpha$ and $\tilde{b}_{\alpha}(J,J')$ are 
defined in the first and the second equations,
respectively. 
Clearly, this NIM-rep is in general not factorizable, since the labels
$\alpha, \beta \in \V$ of the Cardy states are not related with
those of $G$ and $H$. 
As we shall show in the following,
we can construct these \textit{unfactorizable} NIM-reps 
by applying the methods presented in Section \ref{sec:NIMconstruction}
to the $G \times H$ theory.

\subsection{Example: $SU(2) \times  SU(2)$}
\label{sec:tensorex}

In this subsection,
we take a concrete example $G = H = SU(2)$ to
show that the methods in Section \ref{sec:NIMconstruction}
yield non-trivial NIM-reps of the $G \times H$ theory.
This simple example is enough for the illustration of our method.
The generalization to the other cases is straightforward. 

The set of all the integrable representations
of the affine Lie algebra $su(2)_k$ takes the form
\beq
 P_+^k(su(2)) = \{(\mu) | \mu = 0,1,2,\dots,k \} ,
\eeq
where $\mu$ is the Dynkin label of the representation. 
The modular transformation $S$-matrix reads
\beq
   S^{(k)}_{\mu \nu}
     =\sqrt{\frac{2}{k+2}}
      \sin\biggl(\frac{\pi}{k+2}(\mu+1)(\nu+1)\biggr) .
\label{eq:su2ksmatrix}
\eeq
The simple current group is generated by $J = (k)$ and
isomorphic to $\Z_2$,
\beq
  \G(su(2)_k) = \{1 = (0), J = (k)\} \cong \Z_2 .
\eeq
$J$ acts on the $S$ matrix as
\beq
  S^{(k)}_{J\mu\, \nu} = S^{(k)}_{k-\mu \, \nu}
  = S^{(k)}_{\mu \nu} (-1)^\nu . 
\eeq
The $SU(2)_k \times SU(2)_l$ theory is
obtained by taking the product of two factors.
The spectrum $\I$ and
the simple current group $\G$ of the theory are given by
\bea
  \I &= \{(\mu, \nu) | \mu = 0,1,\dots,k; \, \nu = 0,1,\dots,l \} , 
  \label{eq:su2tensorspec} \\
  \G &= \{(1,1), (J,1), (1,J'), (J,J') \} \cong \Z_2 \times \Z_2 .
  \label{eq:su2tensorsc}
\eea

\subsubsection{NIM-rep from simple currents}
\label{sec:tensorexsc}

\begin{figure}[tb]
\begin{center}
\includegraphics[width=7cm]{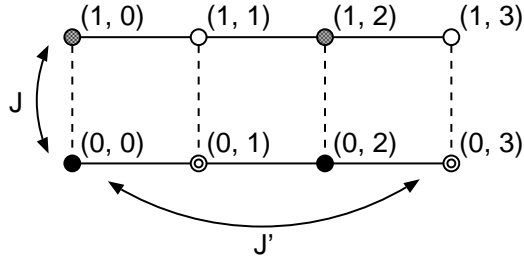}
\end{center}
\caption{The graph for the regular NIM-rep of $SU(2)_1 \times SU(2)_3$.
Each vertex corresponds to the Cardy state.
The dashed (solid) lines stand for the fusion with $(1,0)$ ($(0,1)$).
The arrows show the action of the simple currents
$J = (1,0), J'=(0,3)$ on the Cardy states.
Black, grey, circled and white
vertices express $(\tilde{b}(J,1), \tilde{b}(1,J'))=
(1,1), (-1,1), (1,-1)$ and $(-1,-1)$, respectively.
(see eq.\eqref{eq:scontensorspec}) }
\label{fig:13}
\end{figure}

\begin{figure}[t]
\begin{center}
\includegraphics[width=7cm]{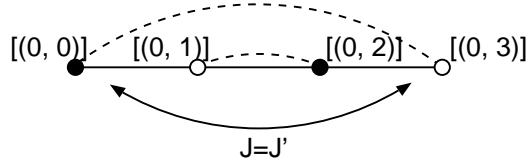}
\end{center}
\caption{The simple current NIM-rep \eqref{eq:scNIMsu2_31}
of $SU(2)_1 \times SU(2)_3$.
The dashed (solid) lines stand for the fusion with $(1,0)$ ($(0,1)$).
The arrows show the action of the simple currents
$J = J'$. Black (white) vertices express
$\tilde{b}(J,J') = 1 \, (-1)$.}
\label{fig:13orb}
\end{figure}

The simple current group \eqref{eq:su2tensorsc} has three non-trivial
subgroups besides the identity and $\G$ itself.
Two of them are the simple current groups of each factor and
not interesting from the point of view of the product theory.
The remaining one is generated by $(J,J')$ and isomorphic to $\Z_2$,
which we denote by $H$,
\beq
  H = \{(1,1), (J,J')\} \cong \Z_2 . 
\eeq
As an example of the simple current NIM-rep,
we consider the $SU(2)_1 \times SU(2)_3$ theory and apply
the method in Section \ref{sec:NIMsc} with the above $H$. 
Our starting point is the regular NIM-rep of
the $SU(2)_1 \times SU(2)_3$ theory. 
The diagonalization matrix is given by the $S$ matrix
\beq
  \psi_{(\alpha,\beta)}{}^{(\mu,\nu)} = 
  S^{(1)}_{\alpha \mu} S^{(3)}_{\beta \nu} =
  \frac{1}{\sqrt{2}} (-1)^{\alpha \mu} S^{(3)}_{\beta \nu}  , \quad
  (\alpha,\beta),  (\mu,\nu) \in \I ,
\eeq
where the spectrum $\I$ reads
\beq
  \I = \{(\mu,\nu)| \mu = 0,1; \nu = 0,1,2,3\} .
\eeq
We have eight Cardy states (see Fig.\ref{fig:13}). 
The group $H \subset \G$ acts on $\psi$ as
\beq
  \psi_{(J \alpha,J' \beta)}{}^{(\mu,\nu)} =
  \psi_{(1 - \alpha, 3 - \beta)}{}^{(\mu,\nu)} =
  \psi_{(\alpha,\beta)}{}^{(\mu,\nu)} (-1)^{\mu + \nu} . 
\eeq
Since there is no fixed point in this action, we can apply
the method in Section \ref{sec:NIMsc} to obtain a new NIM-rep
of the product theory.
Namely, the formula \eqref{eq:scNIM} yields a NIM-rep 
(see Fig.\ref{fig:13orb})
\beq
 \tilde{\psi}_{[(0,\alpha)]}{}^{(\mu,\nu)} = 
 \sqrt{2} \psi_{(0,\alpha)}{}^{(\mu,\nu)} =
 S^{(3)}_{\alpha \nu}  , \quad
 [(0,\alpha)] \in \tilde{\V} , \, (\mu,\nu) \in \tilde{\E} .
 \label{eq:scNIMsu2_31}
\eeq
Here the set $\tilde{\V}$ of the labels of the Cardy states
and the set $\tilde{\E}$ of the labels of the Ishibashi states
are given by
\bea
  \tilde{\V} &= \{ [(0,\alpha)] | \alpha = 0,1,2,3 \} , 
  \label{eq:scNIM13V} \\
  \tilde{\E} &= \{(\mu,\nu) \in \I \, |\, (-1)^{\mu + \nu} = 1 \} 
  = \{(0,0),(1,1),(0,2),(1,3) \} . 
  \label{eq:scNIM13E}
\eea
The action of the simple currents on the Cardy states
has a non-trivial stabilizer
\beq
  \stab{\tilde{\V}} = \{(1,1), (J,J') \} = H \subset \G .
\eeq
The automorphism group $\auto{\tilde{\V}}$ therefore reads
\beq
  \auto{\tilde{\V}} = \G / \stab{\tilde{\V}} \cong \Z_2 . 
\eeq
On the other hand, the automorphism group $\auto{\tilde{\E}}$ takes the form
\beq
  \auto{\tilde{\E}} = \{(1,1), (J,J') \} = H  \cong \Z_2 .  
\eeq
We show the action of these groups in Fig.\ref{fig:13orb}.

One can see that the diagonalization matrix \eqref{eq:scNIMsu2_31}
can not be factorized into the $SU(2)_1$ and the $SU(2)_3$ parts;
the sets $\tilde{\V}$ and $\tilde{\E}$ are not of the factorized form.
This NIM-rep corresponds to the $\Z_2$ orbifold of
$SU(2)_1 \times SU(2)_3$. 
Actually, the partition function of the orbifold can be obtained by
the standard construction and takes the form
\beq
  Z = \abs{\chi_{(0,0)} + \chi_{(1,3)}}^2 
    + \abs{\chi_{(1,1)} + \chi_{(0,2)}}^2 .
  \label{eq:scZsu2_31}
\eeq
The spectrum of the orbifold matches to the set $\tilde{\E}$
obtained above. 

\subsubsection{NIM-rep from conformal embedding}
\label{sec:tensorexembed}

\begin{table}
\renewcommand{\arraystretch}{1.5}
\begin{center}
\begin{tabular}{cccc}
$k$  &  $l$  &  $g$    & $c$  \\ \hline
1    &  3    &  $G_{2,1}$  & $\frac{14}{5}$ \\
2    &  2    &  $A_{3,1}$  & $3$  \\
3    &  8    &  $C_{3,1}$  & $\frac{21}{5}$  \\
2    &  10   &  $D_{4,1}$  & $4$ \\
6    &  6    &  $B_{4,1}$  & $\frac{9}{2}$  \\
8    &  28   &  $F_{4,1}$  & $\frac{26}{5}$ \\
10   &  10   &  $D_{5,1}$  & $5$ 
\end{tabular}
\end{center}
\caption{Conformal embeddings of $su(2)_k \oplus su(2)_l$ into
an affine Lie algebra $g$. 
The subscript $1$ means that the level of the algebra is $1$.
$c$ is the central charge of the embedding.} 
\label{tab:embedding}
\end{table}

The method of conformal embedding also gives us non-trivial
examples of the NIM-reps in tensor product theories. 
The possible embeddings for $su(2)_k \oplus su(2)_l$ is summarized
in Table \ref{tab:embedding} \cite{Stanev}. 
Note that the level of the algebra $g$ is equal to $1$ 
for all the conformal embeddings. 
Among them, we consider the case of $k \ne l$. 
This is because, in the next section, we use the resulting NIM-rep 
in the construction of the NIM-rep
of the diagonal coset theory $SU(2)_{l-k} \times SU(2)_k / SU(2)_l$.
For the coset to be non-trivial, we need $k < l$. 
\footnote{%
The NIM-rep for the case of $k=l$ can be used in
$SU(2)_k \times SU(2)_l / SU(2)_{k+l}$. 
The resulting NIM-rep of the coset theory is, however, 
of the factorized type. 
}
In the following, we consider two examples with $k \ne l$:
$su(2)_1 \oplus su(2)_3 \subset G_{2,1}$ and
$su(2)_3 \oplus su(2)_8 \subset C_{3,1}$. 

\subsubsection*{%
\underline{%
$su(2)_1 \oplus su(2)_3 \subset G_{2,1}$}}

There are two integrable representations in $G_2$ at level $1$,
\beq
  \I = P_+^{1}(G_2) = \{(0,0), (0,1) \} .
  \label{eq:g2spec}
\eeq
In our convention, the short root is $\alpha_2$.
The modular transformation $S$-matrix for $G_{2,1}$ reads
\beq
   S= \frac{2}{\sqrt{5}} 
      \begin{pmatrix}
        \sin \frac{\pi}{5}  &  \sin \frac{2\pi}{5} \\[2\jot]
        \sin \frac{2\pi}{5} & -\sin \frac{\pi}{5} 
      \end{pmatrix} ,
\label{eq:g2smatrix}
\eeq
where the rows and columns are ordered as in eq.\eqref{eq:g2spec}.
These representations branch to those of $su(2)_1 \oplus su(2)_3$ as
follows,
\beq
\renewcommand{\arraystretch}{1.5}
\begin{array}{ll}
 h=0:            & (0,0) \mapsto (0,0) \oplus (1,3) ,\\ 
 h=\frac{2}{5}:  & (0,1) \mapsto (1,1) \oplus (0,2) .
\label{eq:g2toa1a1}
\end{array}
\eeq
Here $h$ is the conformal dimension
of the representation of $G_{2,1}$. 

Let 
$\ket{\alpha} = \sum_{\mu \in \I} S_{\alpha \mu} \dket{\mu}\,
(\alpha \in \{(0,0),(0,1)\})$
be the regular Cardy states of $G_{2,1}$.
Using the branching rule \eqref{eq:g2toa1a1}
and the formula \eqref{eq:Etype}, we can regard these 
boundary states as those of the $SU(2)_1 \times SU(2)_3$ theory.
The spectrum of the Ishibashi states (see eq.\eqref{eq:embedspec})
can be read off from the branching rule,
\beq
  \E_{(1,3)} = \{(0,0),(1,1),(0,2),(1,3) \} . 
  \label{eq:g2embedspec}
\eeq
In terms of these states, the $G_{2,1}$ Cardy state can be written
as a four-dimensional (row) vector $\psi_\alpha$
(see eq.\eqref{eq:embedpsi})
\beq
  \begin{pmatrix} \psi_{(0,0)} \\[2\jot]
                  \psi_{(0,1)} \end{pmatrix}
   = \sqrt{\frac{2}{5}}
     \begin{pmatrix}
          \sin \frac{\pi}{5}  & \sin \frac{2\pi}{5} &
          \sin \frac{2\pi}{5} & \sin \frac{\pi}{5}  \\[2\jot]
          \sin \frac{2\pi}{5} & -\sin \frac{\pi}{5} &
          \sin -\frac{\pi}{5} &  \sin \frac{2\pi}{5}  
     \end{pmatrix} ,
\eeq
where the columns are ordered as in eq.\eqref{eq:g2embedspec}.
Since $\abs{\E_{(1,3)}} = 4$, there are two states missing.
We can obtain these two states by the boundary state generating
technique in Section \ref{sec:embedding}.
(see also Appendix \ref{sec:E6}.)
The resulting boundary states are written as follows,
\beq
  \begin{pmatrix} \psi_0 = \psi_{(0,0)} \\[2\jot]
                  \psi_1 \\[2\jot]
                  \psi_2 = \psi_{(0,1)} \\[2\jot]
                  \psi_3 \end{pmatrix}
   = \sqrt{\frac{2}{5}}
     \begin{pmatrix}
          \sin \frac{\pi}{5}  & \sin \frac{2\pi}{5} &
          \sin \frac{2\pi}{5} & \sin \frac{\pi}{5}  \\[2\jot]
          \sin \frac{2\pi}{5} & \sin \frac{\pi}{5} &
          -\sin \frac{\pi}{5} & -\sin \frac{2\pi}{5} \\[2\jot]
          \sin \frac{2\pi}{5} & -\sin \frac{\pi}{5} &
          -\sin \frac{\pi}{5} &  \sin \frac{2\pi}{5} \\[2\jot] 
          \sin \frac{\pi}{5}  & -\sin \frac{2\pi}{5} &
          \sin \frac{2\pi}{5} & -\sin \frac{\pi}{5}  
     \end{pmatrix} .
\label{eq:g2embedpsi}
\eeq
We obtain four Cardy states from the conformal embedding
$su(2)_1 \oplus su(2)_3 \subset G_{2,1}$.
As is seen from eq.\eqref{eq:g2embedspec}, this NIM-rep
is not factorizable. 

The spectrum \eqref{eq:g2embedspec} is the same as that for
the simple current NIM-rep \eqref{eq:scNIM13E}.
Actually, the NIM-rep obtained above is nothing but that obtained
from the simple current.
This is no coincidence.
The partition function obtained from the affine branching rule
\eqref{eq:g2toa1a1} reads
\beq
  Z = \abs{\chi^{G_2}_{(0,0)}}^2 + \abs{\chi^{G_2}_{(0,1)}}^2
    = \abs{\chi_{(0,0)} + \chi_{(1,3)}}^2 
    + \abs{\chi_{(1,1)} + \chi_{(0,2)}}^2 ,
\eeq
which is exactly that for the orbifold \eqref{eq:scZsu2_31}.

\subsubsection*{%
\underline{%
$su(2)_3 \oplus su(2)_8 \subset C_{3,1}$}}

The construction for the case of 
$su(2)_3 \oplus su(2)_8 \subset C_{3,1}$ proceeds exactly the same way
as above. 
There are four integrable representations in $C_3$ at level $1$,
\beq
  \I = P_+^{1}(C_3) = \{(0,0,0), (1,0,0), (0,1,0), (0,0,1) \} .
  \label{eq:c3spec}
\eeq
In our convention, the long root is $\alpha_3$.
The modular transformation $S$-matrix for $C_3$ at level $1$ 
is given by that for $su(2)_3$,
\beq
   S = \sqrt{\frac{2}{5}}
     \begin{pmatrix}
          \sin \frac{\pi}{5}  & \sin \frac{2\pi}{5} &
          \sin \frac{2\pi}{5} & \sin \frac{\pi}{5}  \\[2\jot]
          \sin \frac{2\pi}{5} & \sin \frac{\pi}{5} &
          -\sin \frac{\pi}{5} & -\sin \frac{2\pi}{5} \\[2\jot]
          \sin \frac{2\pi}{5} & -\sin \frac{\pi}{5} &
          -\sin \frac{\pi}{5} &  \sin \frac{2\pi}{5} \\[2\jot] 
          \sin \frac{\pi}{5}  & -\sin \frac{2\pi}{5} &
          \sin \frac{2\pi}{5} & -\sin \frac{\pi}{5}  
     \end{pmatrix} , 
\label{eq:sp6smatrix}
\eeq
where the rows and columns are ordered as in eq.\eqref{eq:c3spec}. 
The branching rule \eqref{eq:branching}
is found by comparing the conformal dimension,
\beq
\renewcommand{\arraystretch}{1.5}
\left.
\begin{array}{lcl}
 h=0           &: &  (0,0,0)\mapsto(0,0)\oplus (2,4) \oplus (0,8), \\
 h=\frac{7}{20}&: &  (1,0,0)\mapsto(1,2)\oplus (3,4) \oplus (1,6), \\
 h=\frac{3}{5} &: &  (0,1,0)\mapsto(2,2)\oplus (0,4) \oplus (2,6), \\
 h=\frac{3}{4} &: &  (0,0,1)\mapsto(3,0)\oplus (1,4) \oplus (3,8), 
\end{array}
\label{eq:c3toa1a1}
\right.
\eeq
where $(\mu,\nu)$ in the r.h.s. is the representation 
of the $SU(2)_3 \times SU(2)_8$ theory. 
$h$ is the conformal dimension of the representation of $C_{3,1}$.

\begin{figure}[tb]
\begin{center}
\includegraphics[width=8cm]{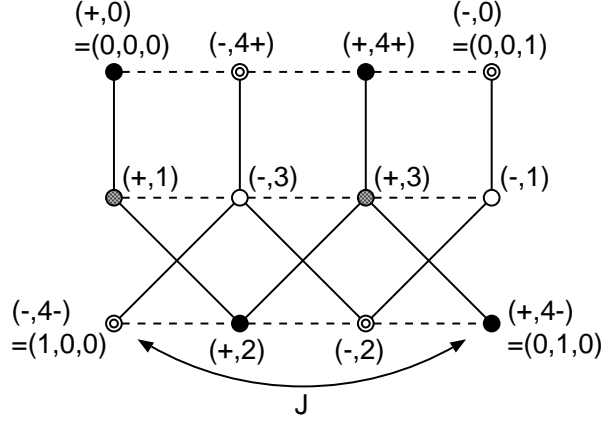}
\caption{The NIM-rep graph of the $SU(2)_3\times SU(2)_8$ theory
from the conformal embedding $su(2)_3 \oplus su(2)_8 \subset C_{3,1}$.
The dashed (solid) lines stand for the fusion with $(1,0)\, ((0,1))$.
The arrow shows the action of the simple currents
$J = (3,0)$ on the Cardy states.
$J' = (0,8)$ belongs to the stabilizer. 
Black, circled, grey and white
vertices express $(\tilde{b}(J,1), \tilde{b}(1,J'))=
(1,1), (-1,1), (1,-1)$ and $(-1,-1)$, respectively.
The vertices connected by a dashed (solid) line
form a $A_4$ ($D_6$) Dynkin diagram. 
}
\label{fig:38}
\end{center}
\end{figure}

From \eqref{eq:c3toa1a1}, one obtains 
the set of Ishibashi states \eqref{eq:embedspec}
\bes
  \E_{(3,8)} =  \{&(0,0), (3,0), (2,2), (1,2), (2,6), (1,6), \\
                          &(0,8), (3,8), (0,4), (3,4), (2,4), (1,4) \}.
\label{eq:38spec}
\ees
The formula \eqref{eq:Etype} gives four boundary states labeled by
$(\lambda_1,\lambda_2,\lambda_3) \in \I$. 
Since $\abs{\E_{(3,8)}}=12$,
there are eight states missing. 
The boundary states generating technique again enables us to construct them.
The missing states are generated from four regular Cardy states of $C_{3,1}$
by taking the fusion with $(1,0)$ and $(0,1)$. 
We show the result in Fig.\ref{fig:38}.
One can see that twelve states are organized in two different ways:
three sets of four states connected by $(1,0)$ and
two sets of six states by $(0,1)$.
The states connected by $(1,0)$ realize the regular NIM-rep of $su(2)_3$,
while those connected by $(0,1)$ are two copies of the $D_6$ NIM-rep
of $su(2)_8$.
The resulting NIM-rep is not of the factorized form, 
rather two kinds of diagrams are folded into one graph
(\textit{cf} Fig.\ref{fig:13}).
It is therefore natural to introduce the label
$(\pm,i)$,~$i=0,1,2,3,4_{\pm},$ 
for the boundary states
\bes
  \V_{(3,8)}=\{&(+,0), (-,0), (+,1), (-,1), (+,2), (-,2), \\
               &(+,3), (-,3), (+,4_+), (-,4_+), (+,4_-), (-,4_-) \}. 
\label{eq:38boundary}
\ees
The first entry distinguishes two copies of $D_6$. 
The sign $\pm$ stands for the eigenvalue $\tilde{b}(J,1)$.
The second entry labels the vertices of $D_6$;
$0$ for the end point of the long leg and $4_\pm$ for two short legs.
In this labeling, the regular Cardy states
$\{(0,0,0), (1,0,0), (0,1,0), (0,0,1)\}$ of $C_{3,1}$ 
correspond to
$\{(+,0), (-,4_-), (+,4_-), (-,0)\}$, respectively.

The action of the simple currents on the Cardy states
has a non-trivial stabilizer
\beq
  \stab{\V} = \{(1,1), (1,J') \} \subset \G .
\eeq
The automorphism group $\auto{\V}$ therefore reads
\beq
  \auto{\V} = \G / \stab{\V} 
  = \{(1,1),(J,1) \} \cong \Z_2 ,
\eeq
while the automorphism group $\auto{\E}$ is $\G$ itself.
We show the action of these groups in Fig.\ref{fig:38}.

We give the explicit form of the diagonalization matrix
$\psi^{(3,8)}{}_{\alpha}{}^{(\mu,\nu)}$, 
\begin{subequations}
\label{eq:38NIMrep}
\beq
  \psi^{(3,8)} = \frac{1}{\sqrt{5}}
      \begin{pmatrix}
      \sqrt{2}c_2 K   &  \sqrt{2}c_1 K
    & \sqrt{2}c_1 K   &  \sqrt{2}c_2 K
    &      K                 &      K        \\[\jot]
      \sqrt{2}s_1 K   &  \sqrt{2}s_2 K
    &-\sqrt{2}s_2 K   & -\sqrt{2}s_1 K
    &        0                    &         0          \\[\jot]
      \sqrt{2}c_1 K   &  \sqrt{2}c_2 K
    & \sqrt{2}c_2 K   &  \sqrt{2}c_1 K
    &     -K                 &     -K        \\[\jot]
      \sqrt{2}s_2 K   & -\sqrt{2}s_1 K
    & \sqrt{2}s_1 K   & -\sqrt{2}s_2 K
    &        0                    &         0          \\[\jot]
      \frac{1}{\sqrt{2}}  K  & -\frac{1}{\sqrt{2}} K
    &-\frac{1}{\sqrt{2}}  K  &  \frac{1}{\sqrt{2}} K
    &    2c_1 K       &  -2c_2 K    \\[\jot]
      \frac{1}{\sqrt{2}}  K  & -\frac{1}{\sqrt{2}} K
    &-\frac{1}{\sqrt{2}}  K  &  \frac{1}{\sqrt{2}} K
    &   -2c_2 K       &    2c_1 K 
      \end{pmatrix} ,
\eeq
where 
\bes
    &K = \frac{1}{\sqrt{2}}
         \begin{pmatrix}
            1  &  1        \\
            1  & -1
         \end{pmatrix} ,    \\
   & c_1 = \cos \frac{\pi}{5}  , \quad
     c_2 = \cos \frac{2\pi}{5} , \quad
     s_1 = \sin \frac{\pi}{5}  , \quad 
     s_2 = \sin \frac{2\pi}{5} . 
\ees
\end{subequations}
The rows $\alpha$ are ordered as in eq.\eqref{eq:38boundary}
whereas the columns $(\mu, \nu)$ are ordered
as in eq.\eqref{eq:38spec}.


\section{Boundary states in coset theories}
\label{sec:coset}

In this section, we develop a method to construct
NIM-reps in coset conformal field theories.
Our strategy is to use the relation of the $G/H$ theory 
with the tensor product theory $G \times H^c$ \cite{GannonWalton}.
Namely, we give a prescription to construct NIM-reps
of coset theories from those of tensor product theories,
which are discussed in the last section. 
As an example, we apply our method to
the $SU(2)_5 \times SU(2)_3 / SU(2)_8$ theory 
to obtain a NIM-rep which is not factorizable into
the $SU(2)_5 \times SU(2)_3$ and $SU(2)_8$ parts.

\subsection{$G/H$ theories}

The $G/H$ theory is based on an embedding of 
the affine Lie algebra $h$ into $g$.
A representation $\mu$ of $g$ is decomposed by 
representations $\nu$ of $h$ as follows,
\beq
 \mu = \dirsum_{\nu}(\mu;\nu)\nu .
\label{eq:cosetbranching}
\eeq 
The spectrum $\hat{\I}$
of the $G/H$ theory consists of all the possible
combination $(\mu; \nu)$,
\bes
  \hat{\I} = \{(\mu; \nu) | \,
       &\mu \in \I^G, \nu \in \I^H, \\
       &b^G_\mu(J) = b^H_\nu(J'),  
        (J \mu; J' \nu) = (\mu; \nu), \forall (J, J') \in \Gid \} .
\label{eq:cosetprimary}
\ees
Here $\I^G$ $(\I^H)$ is the spectrum of the $G$ $(H)$ theory.
$\Gid$ is the group of the identification currents
corresponding to the common center of $G$ and $H$. 
$b^G_\mu(J)$ ($b^H_\nu(J')$) expresses  the monodromy charge of
the $G$ ($H$) theory. 
The condition $b^G_\mu(J) = b^H_\nu(J')$ is the selection rule
of the branching of representations, 
while the relation $(J \mu; J' \nu) = (\mu; \nu)$ is the so-called
field identification.\footnote{%
We do not consider the maverick cosets \cite{FSS,DJ},
for which additional field identifications are necessary.
}%
In this paper, we restrict ourselves to the case that
all the identification orbit have the same length $N_0 = \abs{\Gid}$, 
\beq
  N_0 = \abs{\{(J \mu, J' \nu) \in \I^G \otimes \I^H |\,
               (J, J') \in \Gid \}}
      = \abs{\Gid} . 
  \label{eq:orbitlength}
\eeq
In particular, there is no fixed point in the field identification
\cite{FSS}
\beq
  (J \mu, J' \nu) \neq (\mu, \nu), \quad 
  \forall (\mu, \nu) \in \I^G \otimes \I^H, \quad
  \forall (J, J') \neq 1 \in \Gid .
\eeq
The character of the coset theory is the branching 
function $\chi_{(\mu;\nu)}$ of the algebra 
embedding $h\subset g$. 
From the branching rule \eqref{eq:cosetbranching}, we find
\beq
  \chi_{\mu}^G = \sum_{\nu} \chi_{(\mu;\nu)} \chi_{\nu}^H .
\eeq
The modular transformation $S$-matrix of the $G/H$ theory reads
\beq
  \hat{S}_{(\mu;\nu)(\mu';\nu')}   
      = N_0 S_{\mu \mu'}^G \overline{{S_{\nu \nu'}^H}}
      = N_0 S_{\mu \mu'}^G S^H_{\nu \bar{\nu'}},  
\label{eq:cosetSmatrix}
\eeq
where $N_0$ is the length of the identification orbit
\eqref{eq:orbitlength} and $S^G$ ($S^H$) is the $S$-matrix of
the $G$ ($H$) theory. 

As is seen from the form of the $S$-matrix \eqref{eq:cosetSmatrix},
it is natural to relate the $G/H$ theory
with the tensor product theory $G \times H^c$. 
Here $H^c$ stands for a theory
whose spectrum and $S$-matrix are given by
$\I^H$ and $\overline{S^H}$, respectively.
The spectrum $\I^c$ and the $S$-matrix $S^c$ of the $G \times H^c$ theory
are therefore given by
\bea
  &\I^c =
  \I^{G \times H^c} = \I^G \otimes \I^H  = \I^{G \times H} \equiv \I, 
  \label{eq:GHcspec}\\
  &S^c_{(\mu, \nu) (\mu', \nu')} =
  S^G_{\mu \mu'} \overline{S^H_{\nu \nu'}} =
  S^G_{\mu \mu'} S^H_{\nu \bar{\nu'}} \equiv
  S_{(\mu, \nu) (\mu', \bar{\nu'})} ,
  \label{eq:GHcS}
\eea
where we denote by $\I$ and $S$ the spectrum and the $S$-matrix
of the $G \times H$ theory, respectively. 
The fusion algebra of the $G \times H^c$ theory is the same as
that of the $G \times H$ theory.
From the Verlinde formula \eqref{eq:Verlinde},
one can calculate the fusion coefficients,
\bes
  \mathcal{N}^c_{(\mu, \nu)(\mu', \nu')}{}^{(\mu'', \nu'')}
  &= \sum_{(\rho, \sigma) \in \I} 
    \frac{%
          S^c_{(\mu, \nu)(\rho, \sigma)}
          S^c_{(\mu', \nu')(\rho, \sigma)} 
          \overline{S^c_{(\mu'', \nu'')(\rho, \sigma)}}}
          {S^c_{(0,0)(\rho, \sigma)}}  \\
  &= \sum_{(\rho, \bar{\sigma}) \in \I} 
    \frac{%
          S_{(\mu, \nu)(\rho, \bar{\sigma})}
          S_{(\mu', \nu')(\rho, \bar{\sigma})} 
          \overline{S_{(\mu'', \nu'')(\rho, \bar{\sigma})}}}
          {S_{(0,0)(\rho, \bar{\sigma})}} \\
  &= \mathcal{N}_{(\mu, \nu)(\mu', \nu')}{}^{(\mu'', \nu'')} . 
  \label{eq:GHcfusion}
\ees
The simple current group of the $G \times H^c$ theory
is therefore the same as that of the $G \times H$ theory,
which we denote by $\G \equiv \G^{G \times H}$. 
The action of the simple current group $\G$ on the $S$-matrix reads
\bes
  S_{(J\mu, J'\nu)(\mu', \nu')} &=
  S_{(\mu, \nu)(\mu', \nu')} b_{(\mu', \nu')}(J, J') , \\
  S^c_{(J\mu, J'\nu)(\mu', \nu')} &=
  S^c_{(\mu, \nu)(\mu', \nu')} b^c_{(\mu', \nu')}(J, J') , \quad
  (J, J') \in \G , 
  \label{eq:sconGHc}
\ees
where $b$ and $b^c$ are defined as
\bes
  b_{(\mu, \nu)}(J, J') &= b^G_\mu(J) b^H_\nu(J') , \\
  b^c_{(\mu, \nu)}(J, J') &= b_{(\mu, \bar{\nu})}(J, J')
  = b^G_\mu(J) \overline{b^H_{\nu}(J')} , \quad
  (J, J') \in \G . 
\label{eq:scchargeonGHc}
\ees

In terms of the $G \times H^c$ theory, the identification current
group $\Gid$ is a subgroup of the simple current group $\G$. 
The field identification
$(J \mu; J' \nu) = (\mu; \nu), (J, J') \in \Gid,$
therefore corresponds to taking the quotient of $\I$ by $\Gid$, 
and the selection rule $b^G_\mu(J) = b^H_\nu(J'), (J, J') \in \Gid,$ 
is the condition of the vanishing monodromy charge
$b^c_{(\mu, \nu)} = 1$ with respect to $\Gid$. 
Hence the definition \eqref{eq:cosetprimary} of the spectrum
can be rewritten as follows,
\bes
  \hat{\I} = \{(\mu; \nu) | \,
       & (\mu, \nu) \in \I, \\
       & b^c_{(\mu, \nu)}(J) = 1,  
         J (\mu, \nu) \sim (\mu, \nu), \forall J \in \Gid \} .
\label{eq:cosetprimary2}
\ees
From now on, we denote the simple current of the tensor product theory
by a single letter such as $J$ instead of a pair of letters $(J, J')$. 
The $S$-matrix \eqref{eq:cosetSmatrix} is also expressed as
\beq
  \hat{S}_{(\mu;\nu)(\mu';\nu')}   
      = N_0 S^c_{(\mu, \nu) (\mu', \nu')} .
\label{eq:cosetSmatrix2}
\eeq
This definition of the $S$-matrix $\hat{S}$ has to be consistent
with the field identification.
This is assured by the selection rule,
\beq
  S^c_{J(\mu, \nu)(\mu', \nu')} 
  = S^c_{(\mu, \nu)(\mu', \nu')} b^c_{(\mu', \nu')}(J) 
  = S^c_{(\mu, \nu)(\mu', \nu')} , \quad
 J \in \Gid \quad (b^c_{(\mu', \nu')}(J) = 1) .
\eeq
Here we used the property \eqref{eq:sconGHc}. 
Next, most importantly, $\hat{S}$ should be unitary,
\bes
  \sum_{(\mu'; \nu') \in \hat{\I}}
  \hat{S}_{(\mu; \nu)(\mu'; \nu')} 
  \overline{\hat{S}_{(\mu''; \nu'')(\mu'; \nu')}} 
  &= \sum_{(\mu'; \nu') \in \hat{\I}} N_0^2 
   S^c_{(\mu, \nu)(\mu', \nu')}
   \overline{S^c_{(\mu'', \nu'')(\mu', \nu')}}  \\
  &= \frac{1}{N_0} \sum_{(\mu', \nu') \in \I}
     \frac{1}{N_0} \sum_{J \in \Gid} b^c_{(\mu', \nu')}(J) \, \\
         &\quad\quad\quad\quad \times N_0^2
           S^c_{(\mu, \nu)(\mu', \nu')}
           \overline{S^c_{(\mu'', \nu'')(\mu', \nu')}}  \\
  &= \sum_{J \in \Gid}  \sum_{(\mu', \nu') \in \I}
           S^c_{J(\mu, \nu)(\mu', \nu')}
           \overline{S^c_{(\mu'', \nu'')(\mu', \nu')}}  \\
  &= \sum_{J \in \Gid}
     \delta_{J(\mu, \nu) (\mu'', \nu'')} \\
  &= \delta_{(\mu; \nu)(\mu''; \nu'')} .
\ees
Here we used our assumption of no fixed points to rewrite the sum
\beq
  \sum_{(\mu'; \nu') \in \hat{\I}} \rightarrow \quad
     \frac{1}{N_0} \sum_{(\mu', \nu') \in \I}
     \frac{1}{N_0} \sum_{J \in \Gid} b^c_{(\mu', \nu')}(J) . 
\eeq
The projection operator introduced above takes account of the selection
rule. 

The fusion coefficients of the coset theory can be calculated
via the Verlinde formula \eqref{eq:Verlinde},
\bes
  \hat{\mathcal{N}}_{(\mu; \nu)(\mu'; \nu')}{}^{(\mu''; \nu'')}
  &= \sum_{(\rho;\sigma) \in \hat{\I}} 
    \frac{%
          \hat{S}_{(\mu; \nu)(\rho; \sigma)}
          \hat{S}_{(\mu'; \nu')(\rho; \sigma)} 
          \overline{\hat{S}_{(\mu''; \nu'')(\rho; \sigma)}}}
          {\hat{S}_{(0; 0)(\rho; \sigma)}}  \\
  &= \frac{1}{N_0} \sum_{(\rho, \sigma) \in \I} 
         \frac{1}{N_0} \sum_{J \in \Gid} b^c_{(\rho, \sigma)}(J) 
         \times N_0^2 
         \frac{%
          S^c_{(\mu, \nu)(\rho, \sigma)}
          S^c_{(\mu', \nu')(\rho, \sigma)} 
          \overline{S^c_{(\mu'', \nu'')(\rho, \sigma)}}}
          {S^c_{(0,0)(\rho, \sigma)}}  \\
  &= \sum_{J \in \Gid} 
     \mathcal{N}_{(\mu, \nu)\, J(\mu', \nu')}{}^{(\mu'', \nu'')} .
\label{eq:coset_fusion}
\ees
Here we used the fact \eqref{eq:GHcfusion} that
the fusion rule of the $G \times H^c$ theory is the same as
that of the $G \times H$ theory.

\subsection{NIM-reps in coset theories}
\label{sec:cosetNIM}

Since the $G/H$ theory is related to
the $G \times H^c$ theory, it is natural
to expect that NIM-reps in the $G/H$ theory
is related to NIM-reps in the $G\times H^c$ theory.
In \cite{Ishikawa}, it was shown that a NIM-rep $\hat{\psi}$
in the $G/H$ theory can be constructed as a product of
NIM-reps in the $G$ theory ($\psi^G$) and the $H$ theory ($\psi^H$), 
\beq
  \hat{\psi} \sim \psi^G \overline{\psi^H}. 
  \label{eq:factorizableNIM}
\eeq
Since $\overline{\psi^H}$ is a NIM-rep of the $H^c$ theory, 
the above relation is considered to be a map from NIM-reps
of the $G \times H^c$ theory to those of the $G/H$ theory. 

In this subsection, we extend the construction in \cite{Ishikawa}
to more general class of NIM-reps.
Namely, we give a map from NIM-reps of the $G \times H^c$ theory
to those of the $G/H$ theory,
without the assumption that
NIM-reps in the $G\times H^c$ theory
are factorizable into the $G$ and $H^c$ theories 
as in eq.\eqref{eq:factorizableNIM}. 
The problem is to find a map from a given NIM-rep
\eqref{eq:tensorNIMrep} in the $G\times H^c$ theory
to a NIM-rep in the $G/H$ theory
\beq
  (\hat{n}_{(\mu; \nu)})_{[\alpha]}{}^{[\beta]}
      = \sum_{(\mu'; \nu') \in \hat{\E}}
        \hat{\psi}_{[\alpha]}{}^{(\mu'; \nu')}
        \frac{\hat{S}_{(\mu; \nu)(\mu'; \nu')}}
             {\hat{S}_{(0; 0)(\mu'; \nu')}}
        \overline{\hat{\psi}_{[\beta]}{}^{(\mu'; \nu')}} , \quad
  [\alpha], [\beta] \in \hat{\V} ,
\label{eq:cosetNIM}
\eeq
with an appropriate choice of the sets $\hat{\V}$
of the labels of the Cardy states and $\hat{\E}$
of the Ishibashi states.

The prototype of our map is eq.\eqref{eq:cosetSmatrix2},
which relates the $S$-matrices (i.e., the regular NIM-reps) of
the $G \times H^c$ and $G/H$ theories. 
This relation suggests that a NIM-rep of the $G/H$ theory
can be written in the form
\beq
  \hat{\psi}_{[\alpha]}{}^{(\mu; \nu)}   
      = N \psi^c_{\alpha}{}^{(\mu, \nu)} . \quad
\label{eq:cosetNIMrep}
\eeq
Here $\psi^c$ is the diagonalization matrix of a NIM-rep
of the $G \times H^c$ theory, and $N$ is an appropriate constant.
The label $[\alpha]$ of the boundary states will be defined shortly.

In order to use the above expression \eqref{eq:cosetNIMrep},
we need a NIM-rep $\psi^c$ of the $G \times H^c$ theory. 
As is shown below,
we can construct $\psi^c$ from a NIM-rep of the $G \times H$ theory.
Suppose that $\psi$ gives a NIM-rep of the $G \times H$ theory
as in eq.\eqref{eq:tensorNIMrep}. 
Then one can show that
the following matrix yields a NIM-rep of the $G \times H^c$ theory,
\beq
  \psi^c_\alpha{}^{(\mu, \nu)} = \psi_\alpha{}^{(\mu, \bar{\nu})} ,\quad
  \alpha \in \V^c = \V, \quad (\mu, \nu) \in \E^c . 
\label{eq:GHtoGHc}
\eeq
Here we introduced the set $\E^c$ of the Ishibashi states
for $\psi^c$, which is determined from $\E$ by taking the charge
conjugation only in the $H$ sector, 
\beq
  \E^c = \{(\mu, \nu) | (\mu, \bar{\nu}) \in \E \}. 
\label{eq:cosetspecc}
\eeq
In general, $\E^c \ne \E$, since $(\mu, \nu) \mapsto (\mu, \bar{\nu})$
is not the charge conjugation of the entire theory.
See Appendix \ref{sec:su3} for an example of $\E^c \ne \E$.
One can confirm that $\psi^c$ gives a NIM-rep of the $G \times H^c$
theory as follows,
\bes
  (n^c_{(\mu, \nu)})_{\alpha}{}^{\beta}
      &= \sum_{(\mu', \nu') \in \E^c}
        \psi^c_{\alpha}{}^{(\mu', \nu')}
        \frac{S^c_{(\mu, \nu)(\mu', \nu')}}
             {S^c_{(0, 0)(\mu', \nu')}}
        \overline{\psi^c_{\beta}{}^{(\mu', \nu')}} \\
  &= \sum_{(\mu', \nu') \in \E^c}
        \psi_{\alpha}{}^{(\mu', \bar{\nu'})}
        \frac{S_{(\mu, \nu)(\mu', \bar{\nu'})}}
             {S_{(0, 0)(\mu', \bar{\nu'})}}
        \overline{\psi_{\beta}{}^{(\mu', \bar{\nu'})}} \\
  &= \sum_{(\mu', \nu') \in \E}
        \psi_{\alpha}{}^{(\mu', \nu')}
        \frac{S_{(\mu, \nu)(\mu', \nu')}}
             {S_{(0, 0)(\mu', \nu')}}
        \overline{\psi_{\beta}{}^{(\mu', \nu')}} \\
  &= (n_{(\mu, \nu)})_\alpha{}^\beta  \quad
  (\alpha, \beta \in \V) , 
\ees
where we used eq.\eqref{eq:tensorNIMrep}. 
Hence $\psi^c$ in eq.\eqref{eq:GHtoGHc} gives the same NIM-rep 
as that for the original $G \times H$ theory. 
The special case of the map \eqref{eq:GHtoGHc} is
the relation \eqref{eq:GHcS} of the $S$-matrices $S^c$ and $S$,
for which $\E^c = \I^c = \I$.

We need the action of the simple currents on $\psi^c$.
The action of the simple currents $J \in \G$ on $\psi$
reads (see eq.\eqref{eq:scontensor}) 
\begin{subequations}
\label{eq:scontensor2}
\bea
     \psi_{J \alpha}{}^{(\mu,\nu)}
    & =  \psi_{\alpha}{}^{(\mu,\nu)}
           b_{(\mu, \nu)}(J) ,&& J \in \auto{\V}, 
\label{eq:scontensorbrane2}                                   \\
  {\psi_{\alpha}}^{J(\mu, \nu)}
     & = \tilde{b}_{\alpha}(J) 
         {\psi_{\alpha}}^{(\mu,\nu)} ,&& J \in \auto{\E},
\label{eq:scontensorspec2}
\eea
\end{subequations}
where $b_{(\mu, \nu)}(J)$ is defined in eq.\eqref{eq:scchargeonGHc}.
From this equation and the definition \eqref{eq:GHtoGHc} of $\psi^c$,
the action on $\psi^c$ follows immediately,
\begin{subequations}
\label{eq:scontensorGHc}
\bea
     {\psi^c_{J \alpha}}^{(\mu,\nu)}
    & =  {\psi^c_{\alpha}}^{(\mu,\nu)}
           b^c_{(\mu, \nu)}(J) ,&& J \in \auto{\V^c}, 
\label{eq:scontensorbraneGHc}   \\
  {\psi^c_{\alpha}}^{J(\mu, \nu)}
     & = \tilde{b}^c_{\alpha}(J) 
         {\psi^c_{\alpha}}^{(\mu,\nu)} ,&& J \in \auto{\E^c},
\label{eq:scontensorspecGHc}
\eea
\end{subequations}
where $b^c_{(\mu, \nu)}(J)$ is also defined in eq.\eqref{eq:scchargeonGHc}.
We introduced two groups, $\auto{\V^c}$ and $\auto{\E^c}$.
$\auto{\E^c}$ is the automorphism group of $\E^c$,
\beq
  \auto{\E^c} = \{ J \in \G \,|\, J: \,\E^c  \rightarrow \E^c \} , 
\eeq
which is in general distinct from $\auto{\E}$ since $\E^c \ne \E$. 
The charge $\tilde{b}^c_\alpha$ is defined so that the above equations,
\eqref{eq:scontensorspec2} and  \eqref{eq:scontensorspecGHc},
are consistent with each other. 
$\auto{\V^c}$ needs some explanation. 
As is seen from the definition \eqref{eq:GHtoGHc} of $\psi^c$, 
the set of the labels of the Cardy states for $\psi^c$ is
the same as that for $\psi$, i.e., $\V^c = \V$. 
However, the action of the simple currents on $\V^c$
differs from that on $\V$, namely $b^c(J) \ne b(J)$. 
We therefore introduce the stabilizer for $\V^c$
with respect to the action \eqref{eq:scontensorbraneGHc}, 
which we denote by $\stabc{\V^c}$,
\beq
  \stabc{\V^c} = \{J \in \G |\, b^c_{(\mu, \nu)}(J) = 1 \,\, 
                        \forall (\mu, \nu) \in \E^c \} .
\label{eq:stabilizerC}
\eeq
The automorphism group $\auto{\V^c}$ is defined as
the quotient of $\G$ by this stabilizer,
\beq
  \auto{\V^c} = \G / \stabc{\V^c} . 
\eeq

We return to eq.\eqref{eq:cosetNIMrep} and show that $\hat{\psi}$
gives a NIM-rep of the coset theory. 
We first specify the set $\hat{\E}$ of the labels of the Ishibashi states
for $\hat{\psi}$. 
From the definition \eqref{eq:cosetNIMrep} of $\hat{\psi}$,
one can read off $\hat{\E}$ in the form
\bes
  \hat{\E} &= \{(\mu; \nu) \in \hat{\I} | \, (\mu, \nu) \in \E^c \}  \\
      &= \{(\mu; \nu) |
                (\mu, \nu) \in \E^c,  \,
      b^c_{(\mu, \nu)}(J) =1 \,(\forall J \in \Gid), 
      J(\mu, \nu) \sim (\mu, \nu) \,(\forall J \in \Gid) \}  .
\label{eq:cosetspectemp}
\ees
This expression for $\hat{\E}$ however contains some redundancies.
First, the selection rule
$b^c_{(\mu, \nu)}(J) =1 \,(\forall J \in \Gid)$ has
some trivial equations
if $\Gid$ has elements common with the stabilizer 
$\stabc{\V^c}$ (see eq.\eqref{eq:stabilizerC}).
The selection rules with respect to
$J \in \Gid \cap \stabc{\V^c}$ are trivially satisfied
for $(\mu, \nu) \in \E^c$. 
The non-trivial selection rules are therefore obtained by
considering the quotient group
\beq
  \Gid(\V^c) = \Gid / (\Gid \cap \stabc{\V^c}) \subset \auto{\V^c} .
\label{eq:Gidbrane}
\eeq
The second redundancy in eq.\eqref{eq:cosetspectemp} arises
in the field identification. 
The spectrum $\E^c$ of the Ishibashi states
differs from the spectrum $\I$ of the $G \times H^c$ theory,
and the action of the identification currents $\Gid$ may not close
on $\E^c$.
We are therefore naturally led to the notion of the group of 
the field identifications restricted on $\E^c$,
\beq
  \Gid(\E^c) = \{ J \in \Gid | \, J : \E^c \rightarrow \E^c \}
  = \Gid \cap \auto{\E^c} \subset \auto{\E^c} . 
  \label{eq:Gidspec}
\eeq
With these groups, we can rewrite the expression
\eqref{eq:cosetspectemp} in the form
\bes
  \hat{\E} = \{(\mu; \nu) |
                (\mu, \nu) \in \E^c,  \,
      &b^c_{(\mu, \nu)}(J) =1 \,(\forall J \in \Gid(\V^c) ), \\
      &J(\mu, \nu) \sim (\mu, \nu) \,(\forall J \in \Gid(\E^c) ) \}  .
\label{eq:cosetspec}
\ees

We turn to the set $\hat{\V}$ of the labels of the Cardy states
in the coset theory.
For $\hat{\psi}$ to be well-defined, eq.\eqref{eq:cosetNIMrep} should
be consistent with the field identification.
Since the field identification on $\E^c$ is described by the group
$\Gid(\E^c)$, the consistency condition can be written as
\beq
  \psi^c_\alpha{}^{J(\mu, \nu)} 
  = \tilde{b}^c_\alpha(J) \psi^c_\alpha{}^{(\mu, \nu)} 
  = \psi^c_\alpha{}^{(\mu, \nu)} , \quad \forall J \in \Gid(\E^c) ,
\eeq
where we used eq.\eqref{eq:scontensorspecGHc}.
Hence, the consistency with the field identification
requires the selection rule
$\tilde{b}^c = 1$ for the boundary states. 
We call this condition the brane selection rule \cite{Ishikawa}. 
We can consider the brane selection rule on $\V^c$ as the `dual' of
the field identification on $\E^c$. 
In order to obtain the spectrum $\hat{\E}$ of the coset theory from
that of the tensor product,
we have to impose one more projection, i.e., the selection rule
$b^c_{(\mu, \nu)}(J) = 1 \, (\forall J \in \Gid(\V^c) )$ 
on $\E^c$ (see eq.\eqref{eq:cosetspec}). 
Hence, we have the relation in $\V^c$ dual to the selection rule on $\E^c$.
This is completely parallel to the case of the simple current NIM-rep
in Section \ref{sec:NIMsc}.
In the present case, we take the orbifold by the identification currents
$\Gid(\V^c)$, which organizes the boundary states $\alpha \in \V^c$
into the orbit $[\alpha] = \{J \alpha |\, J \in \Gid(\V^c) \}$. 
The boundary states $\alpha \in \V^c$ of the $G \times H^c$ theory
are therefore identified by the action of $\Gid(\V^c)$.
We call this the brane identification \cite{Ishikawa}.
The set $\hat{\V}$ of the labels of the Cardy states in the coset
theory is therefore defined as follows,
\beq
  \hat{\V} = \{ [\alpha] \,|\,
                \alpha \in \V^c, \,
      J \alpha \sim \alpha \,(\forall J \in \Gid(\V^c)), \,
      \tilde{b}^c_\alpha(J) = 1, \,(\forall J \in \Gid(\E^c)) \} .
\label{eq:cosetboundary}
\eeq
One can see that
this expression has the structure dual to the definition \eqref{eq:cosetspec}
of $\hat{\E}$. 

We have specified the sets $\hat{\V}$ and $\hat{\E}$,
in which the indices of the rows and the columns of $\hat{\psi}$ take values. 
The remaining task is the check that $\hat{\psi}$ gives a NIM-rep
of the coset theory, in particular, that $\hat{\psi}$ is a unitary
matrix.
For $\hat{\psi}$ to give a NIM-rep, it is necessary 
that $\hat{\psi}$ is a square matrix.
This is exactly the same problem that we encountered in
the construction of the simple current NIM-rep in Section \ref{sec:NIMsc},
and can be solved in the same manner if the action of the simple currents
has no fixed points. 
In the construction of $\hat{\psi}$, we have two sets of simple currents,
$\Gid(\E^c)$ and $\Gid(\V^c)$
(see eqs.\eqref{eq:cosetspec} \eqref{eq:cosetboundary}).
For the action of $\Gid(\E^c)$, there is no fixed points
since $\Gid(\E^c)$ is a subgroup of $\Gid$, for which,
in this paper, we assume the absence of the fixed points. 
For the action of $\Gid(\V^c)$, however, there is no obstruction
for the appearance of the fixed points, and we have in general
fixed points (fixed branes) in the action of $\Gid(\V^c)$ on $\V^c$.
This is analogous to the field identification
fixed points in the action of $\Gid$,
and we call these fixed points the brane identification fixed points
\cite{Ishikawa}. 
In the present analysis, we assume that there is no fixed points
in the actions of both $\Gid(\E^c)$ and $\Gid(\V^c)$. 
Then it can be shown in the same way as the simple current NIM-rep that
$\hat{\psi}$ is a square matrix.
With the assumption of no fixed points,
the check for $\hat{\psi}$ to give a NIM-rep of the coset theory
is straightforward,
\bes
  (\hat{n}_{(\mu; \nu)})_{[\alpha]}{}^{[\beta]}
  &= \sum_{(\mu'; \nu') \in \hat{\E}} 
          \hat{\psi}_{[\alpha]}{}^{(\mu'; \nu')}
          \frac{\hat{S}_{(\mu; \nu)(\mu'; \nu')}}
               {\hat{S}_{(0; 0)(\mu'; \nu')}}
          \overline{\hat{\psi}_{[\beta]}{}^{(\mu'; \nu')}} \\
  &= N^2 \frac{1}{\abs{\Gid(\E^c)}} \sum_{(\mu', \nu') \in \E^c} 
         \frac{1}{\abs{\Gid(\V^c)}} \sum_{J \in \Gid(\V^c)}
                                     b^c_{(\mu', \nu')}(J) \\
  & \quad\quad\quad\quad\quad\quad\quad\quad\quad \times
          \psi^c_{\alpha}{}^{(\mu', \nu')}
          \frac{S^c_{(\mu, \nu)(\mu', \nu')}}
               {S^c_{(0, 0)(\mu', \nu')}}
          \overline{\psi^c_{\beta}{}^{(\mu', \nu')}} \\
  &= \frac{N^2}{\abs{\Gid(\E^c)} \abs{\Gid(\V^c)}} 
         \sum_{J \in \Gid(\V^c)}
         \sum_{(\mu', \nu') \in \E^c} 
          \psi^c_{J \alpha}{}^{(\mu', \nu')}
          \frac{S^c_{(\mu, \nu)(\mu', \nu')}}
               {S^c_{(0, 0)(\mu', \nu')}}
          \overline{\psi^c_{\beta}{}^{(\mu', \nu')}} \\
  &= \frac{N^2}{\abs{\Gid(\E^c)} \abs{\Gid(\V^c)}} 
         \sum_{J \in \Gid(\V^c)}
         (n^c_{(\mu, \nu)})_{J \alpha}{}^\beta .
\ees
From this calculation, $\hat{n}$ is a NIM-rep of the coset theory
if we set
$N = \sqrt{\abs{\Gid(\E^c)} \abs{\Gid(\V^c)}}$. 

To summarize, we obtain a NIM-rep of the $G/H$ theory
from that of the $G \times H^c$ theory, 
whose diagonalization matrix is given by
\begin{subequations}
\label{eq:finalformula}
\beq
  \hat{\psi}_{[\alpha]}{}^{(\mu; \nu)} 
  = \sqrt{\abs{\Gid(\E^c)} \abs{\Gid(\V^c)}}
    \psi^c_\alpha{}^{(\mu, \nu)} , \quad
  [\alpha] \in \hat{\V} , \, 
  (\mu; \nu) \in \hat{\E} . 
\label{eq:finalpsi}
\eeq
The NIM-rep for $\hat{\psi}$ is also written as
\beq
 (\hat{n}_{(\mu; \nu)})_{[\alpha]}{}^{[\beta]} = 
         \sum_{J \in \Gid(\V^c)}
         (n^c_{(\mu, \nu)})_{J \alpha}{}^\beta .
\label{eq:finalNIM}
\eeq
\end{subequations}
Note that $\hat{\psi} \hat{\psi}^\dagger = \hat{n}_{(0;0)}$
is not a unit matrix if there is a fixed point in
the action of $\Gid(\V^c)$.
All the results obtained in \cite{Ishikawa} can be reproduced
from this formula by taking an appropriate factorizable
NIM-reps $n = n^G \otimes n^H$ of the $G \times H$ theory.

\subsection{Example: $SU(2) \times SU(2)/ SU(2)$}

We apply the procedure 
developed in the previous subsection to
the diagonal coset $SU(2)_{l-k} \times SU(2)_k / SU(2)_l$
as an illustrative example. 
As we have discussed in the previous subsection,
we can construct a NIM-rep of this model from 
that of $SU(2)_{l-k} \times SU(2)_k \times SU(2)^c_l$.
Since all the representations of $su(2)_l$ is self conjugate,
we can replace $SU(2)^c_l$ with $SU(2)_l$.
Hence our problem reduces to the construction of the NIM-reps
of  $SU(2)_{l-k} \times SU(2)_k \times SU(2)_l$,
which is discussed in Section \ref{sec:tensor}
(in particular Section \ref{sec:tensorex}). 
Among the NIM-reps of this tensor product theory, 
we consider NIM-reps not factorizable into
$SU(2)_{l-k} \times SU(2)_k$ and $SU(2)_l$.
These correspond to NIM-reps
not factorizable into the numerator and the denominator parts
of the coset, 
and yield NIM-reps not constructed previously.

The simplest example of this class of NIM-reps is given by
the tensor product of the regular NIM-rep of $SU(2)_{l-k}$
and an unfactorizable NIM-rep of $SU(2)_k \times SU(2)_l$.
In the following, we use the results of Section \ref{sec:tensorex}
for $(k, l) = (1,3)$ and $(3,8)$
to construct NIM-reps of the corresponding coset theories.
As we shall show below, the case of $(k,l)=(1,3)$,
i.e., the tricritical Ising model $SU(2)_2 \times SU(2)_1/SU(2)_3$,
yields nothing new. The resulting NIM-rep has
brane identification fixed points, and we obtain a factorizable NIM-rep
after the resolution of the fixed points.
This is consistent with the results \cite{BPPZ} that
all the NIM-reps of the minimal models are factorizable. 
On the other hand, the case of $(3,8)$ yields a new NIM-rep,
which corresponds to the exceptional modular invariant
of $SU(2)_5 \times SU(2)_3/SU(2)_8$ \cite{GannonWalton}.

\subsubsection*{%
\underline{%
$SU(2)_2 \times SU(2)_1/SU(2)_3$}}

We consider the NIM-rep of the form $N^{(2)} \otimes n^{(1,3)}$, 
where $N^{(2)}$ is the regular NIM-rep of $SU(2)_2$
and $n^{(1,3)}$ is an unfactorizable NIM-rep of $SU(2)_1 \times SU(2)_3$
from the conformal embedding $su(2)_1 \oplus su(2)_3 \subset G_{2,1}$
constructed in Section \ref{sec:tensorexembed}
($n^{(1,3)}$ is also constructed as the simple current NIM-rep
in Section \ref{sec:tensorexsc}).
The spectrum $\E = \E^c$ of the tensor product theory reads
\beq
  \E = \I^{(2)} \otimes \E^{(1,3)}
  = \{0, 1, 2 \} \otimes \{(0,0), (1,1), (0,2), (1,3) \} ,
\eeq
where $\I^{(2)}$ and $\E^{(1,3)}$ are the spectra of $N^{(2)}$
and $n^{(1,3)}$, respectively (see eq.\eqref{eq:g2embedspec}). 
The labels $\V$ of the boundary states 
of the tensor product theory is given by
\beq
  \V = \I^{(2)} \otimes \V^{(1,3)}
     = \{0, 1, 2 \} \otimes \{0, 1, 2, 3\} ,
\eeq
where for $\V^{(1,3)}$ we adopt the same notation as in 
eq.\eqref{eq:g2embedpsi}.
The simple current group $\G$ is given by the direct product
\beq
  \G = \{1, J=(2)\} \times \{1, J'=(1)\} \times \{1, J''=(3)\} 
  \cong \Z_2 \times \Z_2 \times \Z_2 . 
\eeq
The stabilizer $\stabc{\V^c}$ reads 
\beq
  \stabc{\V^c} = \stab{\V} = \{(1,1,1),(1,J',J'') \} \cong \Z_2 .
\eeq
Since all the representations are self-conjugate, 
the stabilizer $\stabc{\V^c}$ coincides with $\stab{\V}$.
We therefore write $\stab{\V}$ instead of $\stabc{\V^c}$
in the following.
The automorphism groups $\auto{\E}$ and $\auto{\V}$ therefore
take the form
\bes
  \auto{\E} &= \{1, J \} \times \{(1,1), (J', J'') \} \cong \Z_2 \times \Z_2 ,
  \\
  \auto{\V} &= \G / \stab{\V} \cong \Z_2 \times \Z_2 .
\ees
The field identification is generated by the simple current
$(J,J',J'')$,
\beq
  \Gid = \{1, (J,J',J'') \} \cong \Z_2 .
\eeq
The identification current groups for $\E$ and $\V$ read
\beq
  \Gid(\E) = \Gid \cap \auto{\E} = \Gid , \quad
  \Gid(\V) = \Gid / (\Gid \cap \stab{\V}) = \Gid . 
\eeq
From these data and the definitions
\eqref{eq:cosetspec} and \eqref{eq:cosetboundary} of $\hat{\E}$ and
$\hat{\V}$, we obtain
\bea
  \hat{\E} &= \{(0,0;0), (0,1;1), (0,0;2), (0,1;3) \} , \\
  \hat{\V} &= \{(0,0), (0,2), (2,0), (2,2), (1,1), (1,3) \} / 
               (2 - \alpha, \alpha') \sim (\alpha, \alpha') ,
\eea
where
$(\alpha, \alpha') \,
(\alpha \in \I^{(2)}, \, \alpha' \in \V^{(1,3)})$
stands for the boundary states
$\alpha \otimes \alpha'$ in the tensor product theory. 

One can see that the brane identification has two fixed points
$(\alpha, \alpha') = (1,1), (1,3)$. 
In order to obtain a NIM-rep of the coset theory, we
have to resolve these fixed points by introducing two extra Ishibashi
states. 
The coset theory $SU(2)_2 \times SU(2)_1 / SU(2)_3$ 
has exactly two primary fields besides
those appeared in $\hat{\E}$, namely $(1,0;1)$ and $(1,0;3)$. 
Adding these primary fields to $\hat{\E}$, we can resolve the fixed points
to construct a NIM-rep of the coset theory. 
Since $\hat{\E} = \hat{\I}$, one expects that the resulting NIM-rep
is the regular NIM-rep of the coset theory.
We can show this by the explicit construction of the resolved diagonalization
matrix $\hat{\psi}$, which turns out to be
the $S$-matrix of the coset theory.

\subsubsection*{%
\underline{%
$SU(2)_5 \times SU(2)_3/SU(2)_8$}}

The case of $(k,l) = (3,8)$ can be treated in the same way as 
$(1,3)$. 
We start from the tensor product of the form
$N^{(5)} \otimes n^{(3,8)}$, 
where $N^{(5)}$ is the regular NIM-rep of $SU(2)_5$
and $n^{(3,8)}$ is an unfactorizable NIM-rep of $SU(2)_3 \times SU(2)_8$
from the conformal embedding $su(2)_3 \oplus su(2)_8 \subset C_{3,1}$
constructed in Section \ref{sec:tensorexembed}.
The spectrum $\E = \E^c$ of the tensor product theory reads
\bes
  \E = \I^{(5)} \otimes \E^{(3,8)} 
     = &\{(0), (1), (2), (3), (4), (5) \} \\ 
       &\otimes 
        \{(0,0), (3,0), (2,2), (1,2), (2,6), (1,6),  \\
       &\quad\quad
          (0,8), (3,8), (0,4), (3,4), (2,4), (1,4) \}, 
\ees
where $\I^{(5)}$ and $\E^{(3,8)}$ are the spectra of $N^{(5)}$
and $n^{(3,8)}$, respectively (see eq.\eqref{eq:38spec}). 
The labels $\V$ of the boundary states 
of the tensor product theory is given by
\bes
  \V = \I^{(5)} \otimes \V^{(3,8)} 
     = &\{0, 1, 2, 3, 4, 5 \} \\ 
       &\otimes 
        \{(+,0), (-,0), (+,1), (-,1), (+,2), (-,2), \\
       &\quad\quad
          (+,3), (-,3), (+,4_+), (-,4_+), (+,4_-), (-,4_-) \}, 
\ees
where for $\V^{(3,8)}$ is defined in eq.\eqref{eq:38boundary}.
The stabilizer $\stab{\V}$, the automorphism groups and
the identification current groups take the form
\bea
  \stab{\V} &= \{(1,1,1),(1,1,J'') \} \cong \Z_2 , \\
  \auto{\E} &= \G , \quad
  \auto{\V}  = \G / \stab{\V} \cong \Z_2 \times \Z_2 , \\
  \Gid(\E)  &= \Gid \cap \auto{\E} = \Gid , \quad
  \Gid(\V)   = \Gid / (\Gid \cap \stab{\V}) = \Gid . 
\eea
The brane identification group $\Gid(\V) = \Gid$ 
organizes the states in $\V$ into the orbits
\beq
  [(\alpha; +, \alpha')] =
  \{ (\alpha; +, \alpha'), \, (5-\alpha; -, \alpha') \} .
\eeq
where
$(\alpha; \pm, \alpha') \,
(\alpha \in \I^{(5)}, \, (\pm,\alpha') \in \V^{(3,8)})$
stands for the boundary states
$\alpha \otimes (\pm,\alpha')$ in the tensor product theory. 
Clearly, there is no fixed point in the brane identification. 
From eqs.\eqref{eq:cosetspec} \eqref{eq:cosetboundary},
we obtain
\bea
  \hat{\E} &= 
     \bigcup_{\lambda = 0, 2, 4}
     \{(\lambda,0;0), (\lambda,0;4), (\lambda,0;8), 
       (\lambda,2;2), (\lambda,2;4), (\lambda,2;6) \} , 
  \label{eq:358cosetspec} \\
  \hat{\V} &=
      \{ [(\alpha; +, \alpha')] \, | \,
      \alpha  = 0,2,4; \,
      \alpha' = 0,2,4_\pm \} \cup
      \{ [(\alpha; +, \alpha')] \, | \,
      \alpha  = 1,3,5; \,
      \alpha' = 1,3 \} .
\eea
We have $\abs{\hat{\V}} = \abs{\hat{\E}} = 18$ Cardy states.

\begin{figure}[tb]
\begin{center}
\includegraphics[height=7cm]{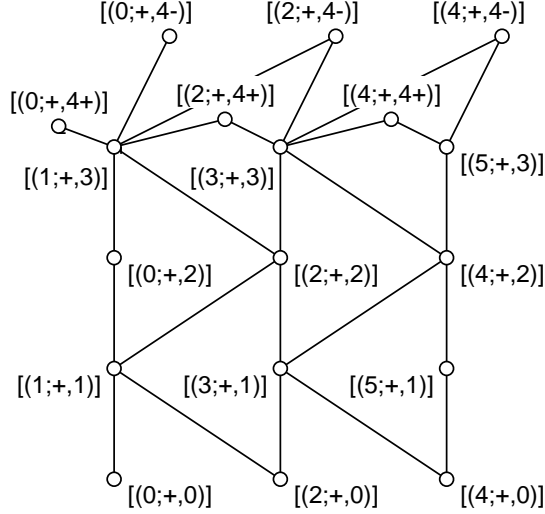}
\end{center}
\caption{The graph 
for an exceptional NIM-rep of
the diagonal coset $SU(2)_5 \times SU(2)_3/SU(2)_8$,
which originates from the conformal embedding
$su(2)_3 \oplus su(2)_8 \subset C_{3,1}$.
There are 18 Cardy states.
The lines stand for the fusion with $(1,0;1)$.}
\label{fig:358e}
\end{figure}

\begin{figure}
\begin{center}
\includegraphics[height=7cm]{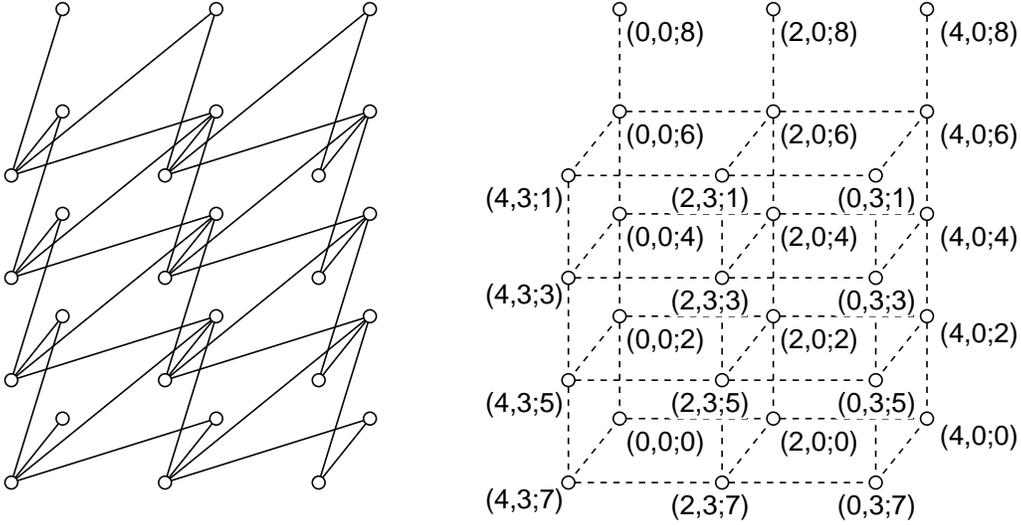}
\end{center}
\caption{The regular NIM-rep of the $SU(2)_5\times SU(2)_3/SU(2)_8$ coset
model. The left figure expresses the fusion with ${(1,0;1)}$,
whereas the right one shows the labels of the corresponding
Cardy states.
The dashed lines are for indicating the relative position
of the vertices. 
Here we give only the half of the Cardy states. 
Total number of the Cardy states
are $\abs{\hat{\I}}= 54 = 27 \times 2$. 
The other half is the same graph with the change that
the second entry of the labels are replaced by $0 \rightarrow 2$
and $3\rightarrow 1$, respectively. }
\label{fig:358r}
\end{figure}

Since there is no fixed point in both $\V$ and $\E$, 
we can use the formula \eqref{eq:finalpsi} to obtain
the diagonalization matrix $\hat{\psi}$,
\beq
 \hat{\psi}_{[(\alpha; +, \alpha')]}{}^{(\lambda,\mu;\nu)}
  = \sqrt{\abs{\Gid(\E)} \abs{\Gid(\V)}}
    \psi_{(\alpha; +, \alpha')}{}^{(\lambda,\mu,\bar{\nu})}
  = 2  S^{(5)}_{\alpha \lambda}
      \psi^{(3,8)}_{(+,\alpha')}{}^{(\mu, \nu)} . 
\label{eq:358psi}
\eeq
Here we denote by $\psi$ the diagonalization matrix 
for the NIM-rep $N^{(5)} \otimes n^{(3,8)}$.
$\psi^{(3,8)}$ is the diagonalization matrix \eqref{eq:38NIMrep}
for $n^{(3,8)}$.
One can observe that
the resulting NIM-rep $\hat{\psi}$ of $SU(2)_5 \times SU(2)_3 / SU(2)_8$
is not factorizable into
the $SU(2)_5 \times SU(2)_3$ and the $SU(2)_8$ parts,
\beq
 \hat{\psi} \not\sim  \psi^{(5,3)}  \psi^{(8)}.
\eeq 
The matrix $\hat{n}_{(\lambda, \mu; \nu)}$ follows from
the formula \eqref{eq:finalNIM},
\bes
  (%
   \hat{n}_{(\lambda, \mu; \nu)}
   )_{[(\alpha;+,\alpha')]}{}^{[(\beta;+,\beta')]} &=
   (n_{(\lambda, \mu, \nu)})_{(\alpha;+,\alpha')}{}^{(\beta;+,\beta')} +
   (n_{(\lambda, \mu, \nu)})_{(5-\alpha;-,\alpha')}{}^{(\beta;+,\beta')} \\
   &=
   (N^{(5)}_\lambda)_{\alpha}{}^{\beta}
   (n^{(3,8)}_{(\mu, \nu)})_{(+, \alpha')}{}^{(+,\beta')} +
   (N^{(5)}_\lambda)_{5-\alpha}{}^{\beta}
   (n^{(3,8)}_{(\mu, \nu)})_{(-, \alpha')}{}^{(+,\beta')} ,
\ees
where
$n_{(\lambda,\mu,\nu)} = N^{(5)}_\lambda \otimes n^{(3,8)}_{(\mu, \nu)}$.
We give the graph for $(\lambda, \mu; \nu) = (1,0;1)$
in Fig.\ref{fig:358e}.
For comparison,
we also give the graph for the regular NIM-rep
of $SU(2)_5 \times SU(2)_3 / SU(2)_8$ in  Fig.\ref{fig:358r}.

One can see that this NIM-rep corresponds to the physical
modular invariant 
$Z_{\mathcal{E}^{(3,8)}}$
of $SU(2)_5\times SU(2)_3/ SU(2)_8$
constructed in \cite{GannonWalton},
\bes
 Z_{\mathcal{E}^{(3,8)}} = \sum_{\mu = 0,2,4} \,\,
     \Bigl[ & \abs{\chi_{(\mu,0;0)}+\chi_{(\mu,0;8)}}^2
        +  \abs{\chi_{(\mu,0;4)}}^2 
        +  \abs{\chi_{(\mu,2;2)}+\chi_{(\mu,2;6)}}^2
        +   \abs{\chi_{(\mu,2;4)}}^2  \\
       &+ (\chi_{(\mu,0;2)}+\chi_{(\mu,0;6)}) \chi_{(\mu,2;4)}^{*}
        +   \chi_{(\mu,2;4)}(\chi_{(\mu,0;2)}+\chi_{(\mu,0;6)})^{*}
                                                               \\
       &+  \chi_{(\mu,0;4)}(\chi_{(\mu,2;0)}+\chi_{(\mu,2;8)})^{*} 
        + (\chi_{(\mu,2;0)}+\chi_{(\mu,2;8)})\chi_{(\mu,0;4)}^{*} \Bigr] ,
\ees
where $\chi_{(\lambda,\mu;\nu)}$ are the characters of the coset theory.
The set of the primary fields which appear in the diagonal terms
of $Z_{\mathcal{E}^{(3,8)}}$, i.e., the spectrum of the Ishibashi states,
precisely coincides with $\hat{\E}$ in eq.\eqref{eq:358cosetspec}.
Hence, the NIM-rep defined in eq.\eqref{eq:358psi} 
is associated with this modular invariant and
is considered to be a physical one. 
The reason for the coincidence of the spectra is that
our NIM-rep \eqref{eq:358psi} is constructed
exactly in the same way as 
the modular invariant $Z_{\mathcal{E}^{(3,8)}}$.
This modular invariant
follows from that for the tensor product theory
$SU(2)_5 \times SU(2)_3 \times SU(2)_8$,
which is obtained by the conformal embedding
$su(2)_3 \oplus su(2)_8 \subset C_{3,1}$ \cite{GannonWalton}.
\footnote{%
To be precise, 
$Z_{\mathcal{E}^{(3,8)}}$ is obtained by combining
the modular invariant from the conformal embedding
with the action of the simple currents $(J,1,J'')$
in $SU(2)_5 \times SU(2)_3 \times SU(2)_8$.
Although we can do the same thing in the construction of the NIM-rep,
the resulting NIM-rep is exactly the same as that obtained
in eq.\eqref{eq:358psi}. 
}
It is therefore natural that the spectrum of
$Z_{\mathcal{E}^{(3,8)}}$ is the same as $\hat{\E}$,
which is also based on the conformal embedding into $C_{3,1}$.

\section{Summary and Discussion}

In this paper,
we have developed a systematic procedure to construct NIM-reps in coset conformal
field theories.
Based on the relation of the $G/H$ theory with the $G \times H$ theory,
we have shown that
any NIM-rep of the $G \times H$ theory yields
a NIM-rep of the $G/H$ theory.
The action of the simple currents on generic NIM-reps
plays an essential role in our construction.
In particular, 
we have identified the NIM-rep version of the field identification
and the selection rule, which was first observed in \cite{Ishikawa}.
Combined with the conformal embedding of $G \times H$, our method
provides a new class of NIM-reps which cannot be factorized into
the $G$ and the $H$ parts.
As an illustrative example of our procedure,
the diagonal coset $SU(2)_5\times SU(2)_3/SU(2)_8$ has been studied in detail.

Toward the classification of the NIM-reps in coset theories,
the most important question is whether our map from the $G \times H$ theory
to the $G/H$ theory gives all the NIM-reps in the coset theory. 
In other words, we should clarify whether our map is onto or not.
If this is the case, the classification problem for the coset theory
reduces to that for the corresponding tensor product theory.
However,
the existence of the brane identification fixed points makes the problem
difficult to study.
To obtain a NIM-rep of the coset theory, we have to resolve
brane identification fixed points, which appear even if there are no fixed
points in the field identification. 
If there are several ways to resolve the brane identification fixed points,
our map is not onto and we have to solve the problem in two steps.  
We note that our map is not one-to-one.
More than one NIM-reps in the tensor product theory are mapped
to the same NIM-rep in the coset theory,
as we have encountered in several examples in this paper. 

Another important problem is the issue of the physical NIM-reps.
Although our method provides NIM-reps in the coset theory, 
it does not assures the existence of the corresponding modular invariant.
It is necessary to find a criterion for our map to give a physical NIM-rep,
such as that obtained in \cite{GannonWalton} for the modular invariant. 

Our construction makes possible 
a parallel treatment of the $G$ and the $H$ sectors of the $G/H$ theory.
In a sense, one can mix the boundary conditions of two sectors. 
It is therefore interesting to clarify the corresponding boundary
condition and geometrical meaning
in the sigma model approach, namely, the gauged WZW models
\cite{MMS,Gawedzki,ES,Fredenhagen,Kubota,Walton}.

The related issue is the application of our method to
the construction of $D$-branes in non-trivial backgrounds, e.g.,
the Kazama-Suzuki models \cite{KS}. 
Recently, it has been found that there exist exotic D-branes
in simple backgrounds such as $c=1$ CFTs
\cite{GRW,GR,Janik,Tseng}
that break the isometry of the target space but keep the (super) conformal 
invariance on the worldsheet. 
It would be interesting to examine
whether our construction gives the similar branes in
the Kazama-Suzuki models.

\vskip \baselineskip
\noindent
\textbf{Acknowledgement: }
The work of T.T. is supported by the Grant-in-Aid for
Scientific Research on Priority Areas (2) 14046201 
from the Ministry of Education, Culture, Science, Sports and Technology
of Japan.

\appendix
\section{The $E_6$ NIM-rep of $su(2)_{10}$
from conformal embedding}
\label{sec:E6}

In this appendix, we illustrate the method of the conformal embedding
presented in Section \ref{sec:embedding} 
by the example $su(2)_{10} \subset sp(4)_1$.
The resulting NIM-rep is the $E_6$ NIM-rep of the $SU(2)$ WZW model
\cite{DZ,BPPZ}.

We begin with the branching rule of the embedding
$su(2)_{10} \subset sp(4)_1$.
The spectrum $\I$ of the $Sp(4)$ WZW model at level 1 is given by
the set of all the integrable representations of $sp(4) = C_2$ 
at level 1, which are labeled by the horizontal part
of the Dynkin label,
\beq
   \I= P_+^{1}(sp(4)) = \{(0,0),\,(1,0),\,(0,1) \}.
\label{eq:sp4spec}
\eeq
Similarly, the spectrum $\tilde{\I}$ of the $SU(2)$ WZW model at level 10
reads
\beq
   \tilde{\I} = P_+^{10}(su(2)) = 
   \{ (\tilde{\mu}) \,|\, \tilde{\mu}=0, 1, \ldots, 10\}.
\eeq
The branching rule \eqref{eq:branching} can be written as follows,
\begin{alignat}{2}
  \label{eq:sp4branching}
     & h=0   &\quad&:(0,0)\mapsto (0) \dirsum (6),\notag   \\
     & h=5/16&&     :(1,0)\mapsto (3) \dirsum (7),            \\
     & h=1/2 &&     :(0,1)\mapsto (4) \dirsum (10) .\notag
\end{alignat}
Here $h$ is the conformal dimension of each representation in $\I$.
The dimension $\tilde{h}_{\tilde{\mu}}$ of $\tilde{\mu} \in \tilde{\I}$
is given by
\beq
  \tilde{h}_{\tilde{\mu}} = \frac{1}{48} \tilde{\mu}(\tilde{\mu} + 2),
\eeq
and consistent with the branching rule \eqref{eq:sp4branching}.
The modular $S$-matrix of $sp(4)_1$ takes the form
\beq
    S = \frac{1}{2}\begin{pmatrix}
                   1        & \sqrt{2}  &      1      \\
                   \sqrt{2} &     0     &   -\sqrt{2} \\
                   1        & -\sqrt{2} &      1          
                   \end{pmatrix},
\label{eq:sp41smatrix}
\eeq
where the rows and columns are ordered as in \eqref{eq:sp4spec}. 
For $su(2)_{10}$, the formula \eqref{eq:su2ksmatrix} yields
\beq
  \tilde{S}_{\tilde{\mu} \tilde{\nu}} =
  \frac{1}{\sqrt{6}} \sin \Bigl(
    \frac{\pi}{12}(\tilde{\mu} + 1)(\tilde{\nu} + 1) \Bigr) . 
\eeq
One can confirm that the branching rule \eqref{eq:sp4branching}
is compatible with the modular transformations
$S$ and $\tilde{S}$.

The starting point of our construction is the regular NIM-rep
$\psi_{\alpha}{}^\mu = S_{\alpha \mu}$ of $sp(4)_1$.
Hence, at the beginning, we have three Ishibashi states and
three Cardy states labelled by
$\E = \V = \I$.
According to the formula \eqref{eq:Ishibashibranching} and
the branching rule \eqref{eq:sp4branching}, 
the Ishibashi states of $sp(4)_1$ can be decomposed into
those of $su(2)_{10}$ as follows, 
\bes
  \dket{(0,0)} &= a \dket{0} + b \dket{6} , \\
  \dket{(1,0)} &= \frac{1}{\sqrt{2}}( \dket{3} +   \dket{7}) , \\
  \dket{(0,1)} &= b \dket{4} + a \dket{10} , 
  \label{eq:sp4ishibashi}
\ees
where 
\beq
   a = \sqrt{\frac{2 \sin(\pi/12)}{\sqrt{6}}}
    =\sqrt{\frac{3-\sqrt{3}}{6}},\quad \quad
   b = \sqrt{\frac{2 \sin(5\pi/12)}{\sqrt{6}}}
    =\sqrt{\frac{3+\sqrt{3}}{6}}.
\eeq
From this expression, one can read off 
the set $\tilde{\E}$ of the labels of the $su(2)$ Ishibashi states,
\beq
   \tilde{\E} =\{0,\,3,\,4,\,6,\,7,\,10\}.
\label{eq:su2sp4spec}
\eeq
Using eq.\eqref{eq:sp4ishibashi},
the Cardy states of $sp(4)_1$ can be regarded
as the boundary states of $su(2)_{10}$.
Since $\abs{\tilde{\E}} = 6$, the Cardy state $\ket{\alpha}$ is expressed
as a six-dimensional (row) vector $\tilde{\psi}_\alpha$
(see eq.\eqref{eq:embedpsi}),
\beq
  \begin{pmatrix} \tilde{\psi}_{(0,0)} \\
                  \tilde{\psi}_{(0,1)} \\
                  \tilde{\psi}_{(1,0)} \end{pmatrix}
   = \begin{pmatrix}
          a/2 & 1/2 & b/2 & b/2 & 1/2  & a/2  \\
          a/2 & -1/2& b/2 & b/2 &-1/2  & a/2        \\
          a/\sqrt{2}&  0  & -b/\sqrt{2}& b/\sqrt{2} 
                          &   0 &  -a/\sqrt{2}  
     \end{pmatrix} ,
\eeq
where the columns are ordered as in \eqref{eq:su2sp4spec}.

Let us apply the boundary state generating
technique in Section \ref{sec:embedding} to the above states.
Since $\abs{\tilde{\E}} - \abs{\V} = 3$, there are three states
missing. 
We consider the fusion of $\tilde{\psi}_{(0,0)}$
with the generator $(1)\in \tilde{\I}$ of
the fusion algebra.
The fusion of $\tilde{\psi}_{(0,0)}$ with $(1)$ yields
the vector $\tilde{\psi}_{(0,0)}\gamma^{(1)}$,
where the generalized quantum dimension $\gamma^{(1)}$ reads
\beq
  \gamma^{(1)} = 
  \mathrm{diag}\biggl(
  \frac{\tilde{S}_{1\tilde{\mu}}}{\tilde{S}_{0\tilde{\mu}}}
  \biggr)_{\tilde{\mu} \in \tilde{\E}} =
  \mathrm{diag}\bigl(
  \sqrt{6} b^2, 1, \sqrt{6} a^2, -\sqrt{6} a^2, -1, \sqrt{6} b^2 \bigr) .
\eeq
One can easily show that
the length of the vector $\tilde{\psi}_{(0,0)}\gamma^{(1)}$ is 1, 
which means that this vector gives one of the Cardy states.
None of the three vectors in $\V$ coincides
with this vector.
Hence $\tilde{\psi}_{(0,0)}\gamma^{(1)}$ gives a new state
\beq
  \tilde{\psi}_{2} = \tilde{\psi}_{(0,0)}\gamma^{(1)}.
\eeq 
Repeating this process for $\tilde{\psi}_{2}$,
we obtain $\norm{\tilde{\psi}_{2}\gamma^{(1)}}^2 = 2 $,
which means that $\tilde{\psi}_{2}\gamma^{(1)}$ contains
two Cardy states. 
One can easily identify one of the Cardy states with
$\tilde{\psi}_{(0,0)}$, 
since
$(\tilde{\psi}_{2}\gamma^{(1)}, \tilde{\psi}_{(0,0)}) =
\tilde{\psi}_{2}\gamma^{(1)} \tilde{\psi}_{(0,0)}{}^\dagger = 1$.
Subtracting $\tilde{\psi}_{(0,0)}$ from $\tilde{\psi}_{2}\gamma^{(1)}$,
we obtain a new vector
\beq
  \tilde{\psi}_{3} = \tilde{\psi}_{2}\gamma^{(1)}-\tilde{\psi}_{(0,0)}.
\eeq
The fusion of $\tilde{\psi}_3$ also yields a new state
\beq
  \tilde{\psi}_4 =
  \tilde{\psi}_3\gamma^{(1)}-\tilde{\psi}_{2}-\tilde{\psi}_{(1,0)}.
\eeq
The fusion of $\tilde{\psi}_4$ however gives
no new states
\beq
  \tilde{\psi}_{4} \gamma^{(1)} =  
  \tilde{\psi}_{(0,1)} + \tilde{\psi}_3 , 
\eeq
and the generating process terminates here.
We have therefore obtained six mutually independent Cardy states,
which are labelled by the set
\beq
  \tilde{\V}=\{ 1 = (0,0),\, 2,\, 3,\, 4,\,
                 5 = (0,1),\, 6 = (1,0)\}.
\label{eq:su2sp4boundary}
\eeq
The explicit form of the matrix $\tilde{\psi}$ reads
\beq
   \tilde{\psi} = 
   \begin{pmatrix}
     \tilde{\psi}_1 = \tilde{\psi}_{(0,0)} \\
     \tilde{\psi}_2 \\
     \tilde{\psi}_3 \\
     \tilde{\psi}_4 \\
     \tilde{\psi}_5 = \tilde{\psi}_{(0,1)}\\
     \tilde{\psi}_6 = \tilde{\psi}_{(1,0)} \end{pmatrix} =
          \begin{pmatrix}
          a/2 & 1/2 & b/2 & b/2 & 1/2  & a/2  \\
          b/2 & 1/2 & a/2 &-a/2 & -1/2 & -b/2 \\
          b/\sqrt{2}&  0  &-a/\sqrt{2} & -a/\sqrt{2} 
                          &     0      & b/\sqrt{2} \\
          b/2 & -1/2& a/2 &-a/2 & 1/2  & -b/2       \\
          a/2 & -1/2& b/2 & b/2 &-1/2  & a/2        \\
          a/\sqrt{2}&  0  & -b/\sqrt{2}& b/\sqrt{2} 
                          &   0 &  -a/\sqrt{2}  
          \end{pmatrix},
\eeq
which coincides with the diagonalization matrix
of the $E_6$ NIM-rep of $su(2)_{10}$ \cite{DZ,BPPZ}.
\begin{figure}[tb]
\begin{center}
\includegraphics[height=3.5cm]{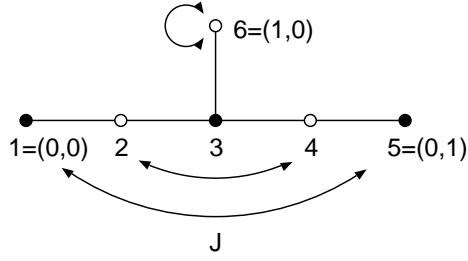}
\end{center}
\caption{The $E_6$ NIM-rep from the conformal embedding
$su(2)_{10} \subset sp(4)_1$. Each vertex corresponds to the Cardy state.
The solid line stands for the fusion with $(1)$. 
The arrows show the action of the simple current $J = (10)$
on the Cardy states.
A black (white) vertex has $\tilde{b} = 1$ ($\tilde{b}= -1$). 
(see eq.\eqref{eq:sconspec}) }
\label{fig:E6}
\end{figure}
We show the result in Fig.\ref{fig:E6} together with the 
action of the simple current \eqref{eq:sconNIM}.

\section{Construction of NIM-reps in the 
$SU(3)_1\times SU(3)_1$ $/SU(3)_2$ model}
\label{sec:su3}

In this appendix, 
we apply the method developed in Section \ref{sec:cosetNIM}
to the diagonal coset $SU(3)_1\times SU(3)_1/SU(3)_2$.
We start from the simple current NIM-reps of
the $SU(3)_1 \times SU(3)_1 \times SU(3)_2$ model
with the sets $\E$ and $\V$.
Since the representations of $su(3)$ are in general not real,
we have to distinguish $\E^c$ with $\E$ (see eq.\eqref{eq:cosetspecc}).
Hence, this model is a simple example for illustrating the meaning of 
the charge conjugation $\E \rightarrow \E^c$,
which plays an essential role in the construction of NIM-reps in the
coset theories. 

The set of the primary fields of the $SU(3)_1\times SU(3)_1\times SU(3)_2$ 
theory is given by
\beq
 \I =  \{(\lambda,\mu,\nu)\,|\, 
               \lambda,\mu \in P_+^1(su(3)); \, 
               \nu \in P_+^2(su(3)) \},
\eeq 
where
\beq
   P_+^1(su(3)) =\{ 00,\, 10,\, 01\}, \quad
   P_+^2(su(3)) =\{ 00,\, 20,\, 02,\, 11,\, 01,\, 10\},
\eeq
and $(\lambda_1 \lambda_2)$ are the Dynkin labels of $su(3)$.
The simple current group of $su(3)_k$ are generated by $(k0)$ 
and isomorphic to $\Z_3$,
\beq
  \G(su(3)_k) = \{1 = (00), J = (k0), J^2 = (0k) \} \cong \Z_3 ,
\eeq
which acts on $P_+^{1}(su(3))$ and $P_+^{2}(su(3))$ as
\bes
  J:  & ~00 \rightarrow 10 \rightarrow 01 \rightarrow 00,       \\
  J: & ~00 \rightarrow 20 \rightarrow 02 \rightarrow 00 ~,~ 
        11 \rightarrow 01 \rightarrow 10 \rightarrow 11 .
\label{eq:su3sc}
\ees
The monodromy charge is given by
\beq
  b_\lambda(J) = \exp\biggl(\frac{2 \pi i}{3}(\lambda_1+2\lambda_2)\biggr).
\label{eq:su3mc}
\eeq

The simple current group $\G$ of $SU(3)_1 \times SU(3)_1 \times SU(3)_2$
is the direct product of each factor,
\beq
  \G = \{(J^l, J^m, J^n) | \, l,m,n = 0,1,2 \}
  \cong \Z_3 \times \Z_3 \times \Z_3 . 
\eeq
As is discussed in Section \ref{sec:NIMsc},
we can construct the simple current NIM-rep of the product theory
for any subgroup $H$ of $\G$. 
Here we take the following $H$,
\beq
   H = \{(1,1,1),\, (1,J,J),\, (1,J^2,J^2)\} \cong \Z_3 . 
\eeq
Since the action of $H$ has no fixed points,
we can use the formula \eqref{eq:scNIM} to obtain the simple current
NIM-rep,
\beq
  \psi_{(\alpha:\beta:\gamma)}{}^{(\lambda, \mu, \nu)}
  = \sqrt{3} S^{(1)}_{\alpha \lambda} 
             S^{(1)}_{\beta \mu}
             S^{(2)}_{\gamma \nu} , \quad
  (\alpha:\beta:\gamma) \in \V, \,\, 
  (\lambda, \mu, \nu) \in \E ,
\eeq
where $S^{(k)}$ is the modular transformation $S$-matrix of $su(3)_k$.
The sets $\V$ and $\E$ are defined as follows,
\bes 
  \V &= \{(\alpha:\beta:\gamma) | \,
          (\alpha, \beta, \gamma) \in \I, 
          (\alpha: J \beta : J \gamma) = (\alpha: \beta: \gamma) \} \\
     &= \{(\alpha: \beta: 00 ), (\alpha: \beta: 11) | \,
           \alpha, \beta \in P_+^{1}(su(3)) \}, \\
  \E &= \{(\lambda,\mu,\nu) \in \I \, | \,
            b_{\mu}(J) b_{\nu}(J) = 1 \} \\
     &= \{(\lambda, 00, 00), (\lambda, 00, 11), 
          (\lambda, 10, 20), (\lambda, 10, 01), \\ 
     &  \quad\quad   
          (\lambda, 01, 02), (\lambda, 01, 10) | \,
           \lambda \in P_+^{1}(su(3)) \}  .
\label{eq:su3tensorspecboundary}
\ees

Next, we use the procedure in Section \ref{sec:cosetNIM} to obtain
the corresponding NIM-rep in the $SU(3)_1\times SU(3)_1$ $/SU(3)_2$
theory. 
Applying the charge conjugation
$\overline{(\lambda_1 \lambda_2)} = (\lambda_2 \lambda_1)$ to $\E$, 
we obtain the set $\E^c$,
\bes
  \E^c &= \{(\lambda, 00, 00), (\lambda, 00, 11), 
            (\lambda, 10, 02), (\lambda, 10, 10), \\ 
       &  \quad\quad   
            (\lambda, 01, 20), (\lambda, 01, 01) | \,
             \lambda \in P_+^{1}(su(3)) \}  .
\ees
Note that $\E^c \neq \E$.
Since the stabilizer of $\V^c$ is $H$,
the automorphism groups for $\E^c$ and $\V^c$ read
\bes
  \auto{\E^c} &= \{(J^n, 1, 1), (J^n, J, J^2), (J^n, J^2, J) | \,
                   n = 0, 1, 2 \} \cong \Z_3 \times \Z_3 , \\
  \auto{\V^c} &= \G / H \cong \Z_3 \times \Z_3 . 
\ees
The identification current group is
\beq
  \Gid = \{(J^n,J^n,J^n) \,|\, n=0,1,2\} \cong \Z_3 ,
\eeq 
whereas the identification groups for $\E^c$ and $\V^c$ are
\bes
  \Gid(\E^c) &= \Gid \cap \auto{\E^c} = \{(1,1,1)\} , \\
  \Gid(\V^c) &= \Gid / (\Gid \cap H) = \Gid .
\ees
The brane identification group $\Gid(\V^c) \cong \Z_3$ acts on $\V^c$ as
\beq
  J:\,
    (\alpha:\beta:\gamma) \rightarrow (J\alpha:J\beta:J\gamma)
    = (J\alpha:\beta:\gamma) .
\eeq  
From eqs.\eqref{eq:cosetspec} \eqref{eq:cosetboundary}, we obtain
\bes
  \hat{\E} &= \{ (00, 00; 00), (00, 00; 11),
                 (00, 10; 10), (00, 10; 02),
                 (00, 01; 01), (00, 01; 20) \} , \\
  \hat{\V} &= \{ [(00:00:00)], [(00:00:11)], [(00:10:00)], \\
           & \quad\quad
                 [(00:10:11)], [(00:01:00)], [(00:01:11)]\}.
\ees
The diagonalization matrix follows from the formula \eqref{eq:finalpsi},
\beq
   \hat{\psi}_{[(\alpha:\beta:\gamma)]}{}^{(\lambda,\mu;\nu)}
  =\sqrt{\abs{\Gid(\E^c)} \abs{\Gid(\V^c)}}
         \psi_{(\alpha:\beta:\gamma)}{}^{(\lambda,\mu,\bar{\nu})}  
  = 3 S^{(1)}_{\alpha\lambda}
       S^{(1)}_{\beta\mu}
       S^{(2)}_{\gamma \bar{\nu}} 
  = \hat{S}_{(\alpha,\beta; \gamma) (\lambda,\mu;\nu)} ,
\label{eq:su3112NIMrep}
\eeq
where $\hat{S}$ is the $S$-matrix of the coset theory.
The resulting NIM-rep is nothing but the regular NIM-rep
of the coset theory and manifestly physical. 
In this coset theory, 
there is another NIM-rep which follows from the twisted boundary condition
of $su(3)$ \cite{Ishikawa}.
The twisted NIM-rep is also constructed 
as the simple current NIM-rep by
taking an appropriate $H$, e.g.,
a subgroup generated by $(1,J,J^2)$.


\end{document}